\documentclass[conference]{IEEEtran}
\usepackage{fancyhdr}
\IEEEoverridecommandlockouts

\usepackage{graphicx} 
\usepackage[normalsize]{subfigure}
\usepackage{epstopdf}
\usepackage{braket}
\usepackage{algorithmic}

\usepackage{amsmath, amssymb} 
\usepackage{xspace}
\usepackage{epsfig}
\usepackage{tikz}
\usepackage{amsmath}
\usepackage{listings}
\lstdefinestyle{bahamas}{
  language=Python,
  basicstyle=\ttfamily\footnotesize,
  keywordstyle=\color{blue}\bfseries,
  stringstyle=\color{red!70!black},
  commentstyle=\color{green!50!black}\itshape,
  identifierstyle=\color{black},
  emph={Bahamas,from_qiskit,from_pennylane,optimize,infer},
  emphstyle=\color{purple!80!black}\bfseries,
  backgroundcolor=\color{gray!8},
  frame=single,
  rulecolor=\color{gray!40},
  framexleftmargin=2pt,
  xleftmargin=6pt,
  xrightmargin=4pt,
  aboveskip=4pt,
  belowskip=4pt,
  columns=flexible,
  keepspaces=true,
}
\usepackage{mathtools}
\usepackage{ellipsis}
\usepackage{textcomp}
\usepackage{hyperref}
\usepackage{booktabs}
\usepackage{caption}
\usepackage{balance}
\usepackage{makecell}

\usepackage[nocompress]{cite}

\usepackage{bm}

\usepackage{tikz}
\usepackage{filecontents}

\usepackage[most]{tcolorbox}

\tcbset{
  enhanced,                
  colback=gray!5!white,    
  colframe=black!80!black, 
  boxrule=0.4pt,           
  arc=4pt,                 
  left=2pt,                
  right=2pt,               
  top=2pt,                 
  bottom=2pt,              
  fonttitle=\bfseries,     
  before skip=6pt,         
  after skip=6pt,          
}

\usepackage{multirow}
\usepackage{hhline}
\usepackage{tabularx}
\newcolumntype{Y}{>{\centering\arraybackslash}X}
\usepackage[inline]{enumitem}
\usepackage{url, comment}
\usepackage{graphicx}

\usepackage{ifthen}
\usepackage{xspace}
\usepackage{epstopdf}
\usepackage{booktabs}

\usepackage{hyperref}

\hypersetup{
  colorlinks=true,      
  linkcolor=blue,       
  citecolor=magenta,    
  filecolor=cyan,       
  urlcolor=red          
}

\usepackage{algorithm}

\newlength\myindent
\newcommand{\sol}[1]{BAHAMAS}

\newcommand{\paraspace}{\vspace{0.05in}}
\newcommand{\parab}[1]{\paraspace\noindent{\bf #1} }

\usepackage{tcolorbox}
\usepackage[framemethod=TikZ]{mdframed}
\mdfdefinestyle{MyFrame}{%
    linecolor=black,
    outerlinewidth=0.5pt,
    innertopmargin=0.5\baselineskip,
    innerbottommargin=0.5\baselineskip,
    backgroundcolor=gray!50!white
}

\usepackage{booktabs}
\usepackage{multirow}
\usepackage{colortbl}

\usepackage{algorithm}
\usepackage{algorithmic}

\definecolor{headergray}{RGB}{210,210,210}
\definecolor{bluegray}{rgb}{0.4, 0.6, 0.8}
\definecolor{pastelblue}{rgb}{0.68, 0.78, 0.81}
\definecolor{palecornflowerblue}{rgb}{0.67, 0.8, 0.94}
\definecolor{paleaqua}{rgb}{0.74, 0.83, 0.9}
\definecolor{lightcornflowerblue}{rgb}{0.6, 0.81, 0.93}
\definecolor{lightblue}{rgb}{0.68, 0.85, 0.9}
\definecolor{cambridgeblue}{rgb}{0.64, 0.76, 0.68}
\definecolor{ashgrey}{rgb}{0.7, 0.75, 0.71}
\definecolor{beaublue}{rgb}{0.74, 0.83, 0.9}
\definecolor{lightcornflowerblue}{rgb}{0.6, 0.81, 0.93}
\definecolor{moonstoneblue}{rgb}{0.45, 0.66, 0.76}
\definecolor{palecornflowerblue}{rgb}{0.67, 0.8, 0.94}
\definecolor{powderblue}{rgb}{0.69, 0.88, 0.9}
\definecolor{blockblue}{RGB}{220,230,242}
\definecolor{blockgreen}{RGB}{221,235,229}
\definecolor{blockyellow}{RGB}{240,232,210}

\definecolor{lightblue}{HTML}{E6F0FF}
\definecolor{lightgreen}{HTML}{E9F7EF}
\definecolor{lightyellow}{HTML}{FFF9E6}
\definecolor{lightpink}{HTML}{FDEDEC}
\definecolor{headergray}{HTML}{F2F2F2} 

\definecolor{headerteal}{RGB}{100,180,180}
\definecolor{lightyellow}{RGB}{255,255,180}
\definecolor{lightlavender}{RGB}{210,190,240}

\definecolor{headerlight}{RGB}{180,220,220}

\definecolor{lightyellow}{HTML}{E6F1FB}
\definecolor{lightlavender}{HTML}{FAEEDA}
\definecolor{lightlavendernew}{HTML}{EDE9FE}

\definecolor{itergray}{RGB}{220, 220, 220}
\definecolor{mapone}{RGB}{198, 235, 210}   
\definecolor{maptwo}{RGB}{198, 218, 240}   
\definecolor{mapthree}{RGB}{255, 223, 186} 

\usepackage{bbding}
\usepackage{pifont}

\usepackage[lastpage,user]{zref}

\usepackage{eso-pic}

\begin{document}

\title{\sol{}: A Control Plane for Optimization\\ and Execution of Variational Quantum Circuits
}

\newcommand{\authcol}[3]{%
  \begin{minipage}[t]{0.3\linewidth}\centering
    {\sublargesize #1}\\[0.0ex]
    {\normalsize #2\\#3}
  \end{minipage}}
\author{%
  \authcol{Amit Samanta}{University of Utah}{Salt Lake City, UT, USA}\hfill
  \authcol{Mohammad Abrarul Hasanat}{University of Utah}{Salt Lake City, UT, USA}\hfill
  \authcol{Jason Ludmir}{Rice University}{Houston, TX, USA}\\[3ex]
  \authcol{Ryan Stutsman}{University of Utah}{Salt Lake City, UT, USA}\hfill
  \authcol{Tirthak Patel}{Rice University}{Houston, TX, USA}\hfill
  \authcol{Rohan Basu Roy}{University of Utah}{Salt Lake City, UT, USA}%
}

\maketitle

\AddToShipoutPictureBG*{%
  \AtTextLowerLeft{%
    \put(0,-24){%
      \parbox{\columnwidth}{\footnotesize This work is accepted for publication at
      the ACM/IEEE International Conference for High Performance Computing,
      Networking, Storage, and Analysis (Supercomputing, SC 2026).}%
    }%
  }%
}

\begin{abstract}

\textit{Variational quantum algorithms (VQAs) suffer from unstable optimization due to temporal noise drift and static qubit mappings that distort gradient signals across iterations. We present \sol{}, an online control framework that stabilizes noise exposure by adaptively selecting physical mappings via consensus-based fidelity estimation, without requiring simulators, offline training, or prior executions. Across real quantum devices, \sol{} improves optimization reliability and supports inference-time retargeting under drift through robust, per-iteration control.}
\end{abstract}

\section{Introduction}
\label{sec:intro}

\begin{figure}[t] 
\centering
\includegraphics[scale = 0.46]{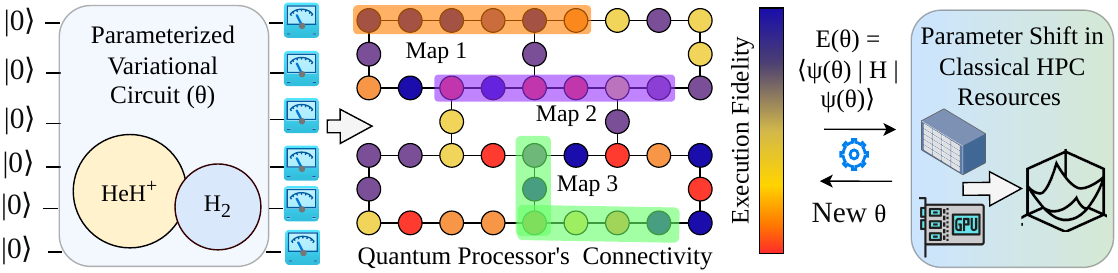}
\vspace{1.5mm}
\caption{\textit{Hybrid classical-quantum optimization loop. The parameterized circuit (ansatz) maps onto the processor via different physical mappings, each with a different fidelity. Cost and gradients are computed classically via parameter-shift, and the parameters are updated for the next iteration.}}
\vspace{-6mm}
\label{fig:background}
\end{figure}

\noindent\textbf{The problem.} Variational quantum algorithms (VQAs) solve problems in quantum chemistry, combinatorial optimization, and machine learning by coupling quantum circuit execution with classical parameter optimization on HPC resources. A parameterized circuit runs on the quantum processor, measurement outcomes yield a cost, and a classical optimizer computes updates over iterations~\cite{cerezo2021variational,
sharma2020noise}. This hybrid loop spans three system layers -- transpilation (compilation), runtime execution on the quantum processor, and parameter update on classical HPC resources, and it is the primary integration point between quantum hardware and classical HPC infrastructure~\cite{long2025hybrid} (Fig.~\ref{fig:background}, more details in Sec.~\ref{sec:background}). The classical side assumes that cost evaluations across iterations are comparable. However, the quantum side often violates this assumption. Before optimization, the transpiler compiles the logical circuit onto a specific subset of physical qubits and does not revisit this decision. Different subsets expose the circuit to different quantum hardware noise, and the same subset produces different noise across iterations as the device drifts. The optimizer does not have visibility into this. When noise changes the relative ranking of parameter configurations, the optimizer receives misleading signals that accumulate and destabilize convergence~\cite{fontana2021evaluating}.

\vspace{2mm}

\noindent\textbf{The challenge.} Variational optimization needs a stable execution regime, \textit{i.e.,} consistent noise conditions across gradient calculations and parameter updates so that the optimizer can make meaningful progress. Quantum hardware does not provide this guarantee. The transpiler makes a one-time qubit assignment~\cite{javadi2024quantum}. The runtime does not track how noise conditions change between iterations~\cite{patel2020experimental}. Also, the optimizer assumes measurements are comparable~\cite{cerezo2021variational}. These three layers operate independently, and the coordination between them for stable execution is missing from current workflows. We observe that fixing a qubit assignment fails because the best one changes over time, and greedy per-iteration switching disrupts noise consistency between updates. Existing techniques that address noise through post-execution error mitigation~\cite{kim2023scalable} or noise-aware compilation~\cite{tannu2019not} do not provide per-iteration control over qubit assignment and parameter update quality jointly across the optimization and inference lifecycle. This problem persists as hardware improves. Error correction reduces gate error rates, but it does not eliminate qubit noise heterogeneity or calibration drift in superconducting qubits~\cite{dangwal2025variational}.

\vspace{2mm}

\noindent\textbf{Our approach.}
This paper presents \sol{}, a control plane for variational quantum circuits on HPC-quantum hybrid systems. A key insight is that what preserves optimization quality is not just low noise at any single iteration but consistent noise exposure across consecutive iterations. \sol{} coordinates qubit assignment and parameter update quality per iteration across the full optimization and inference lifecycle through three components: noise estimation and stabilization, trust-aware parameter update control, and inference-time retargeting on drifted or topologically different backends. The control abstraction maintains two properties at every iteration -- low and stable noise exposure on the cost function, and conditional trust over parameter updates (Sec.~\ref{sec:control_abstraction}). The system operates across all three layers of the stack through online feedback. It does not require any offline training, simulators, or learned models, which makes the control plane robust to backend drift and deployable without external data pipelines. We evaluate \sol{} on current IBM quantum processors across multiple generations and topologies, including early fault-tolerant systems. The control plane remains to be effective in fault-tolerant noise conditions because qubit heterogeneity and calibration drift still persist~\cite{google2025quantum,fang2025caliqec}.

\vspace{2mm}

\noindent\textbf{In summary, we make the following key contributions.}
\begin{enumerate}
\item We identify stable noise exposure across iterations as the property that preserves optimization quality, and define a control abstraction based on noise stability and conditional parameter update trust.

\item We design \sol{}, a control plane across transpilation, execution, and parameter update that manages qubit assignment and trust-aware parameter updates per iteration through online calibration feedback.

\item We show that co-locating candidate qubit assignments on the chip and comparing their outputs provides a practical quality signal for per-iteration selection within realistic quantum execution budgets.

\item We extend the control plane to inference through retargeting on drifted and topologically different backends, without requiring circuit reoptimization.

\item We evaluate \sol{} on real quantum hardware across multiple processor generations. \sol{} improves VQA performance by over 20\% on average, composes with complementary techniques, retains gains under drift and cross-backend transfer, and remains effective as noise levels decrease toward fault-tolerant regimes.
\end{enumerate}

\parab{Open-source artifact.} \sol{} is available at the following link: https://doi.org/10.5281/zenodo.21513003.
\section{Background and Motivation}
\label{sec:background}

\noindent\textbf{Qubits, gates, and noise.}
A quantum computer operates on qubits using single-qubit rotation gates and two-qubit entangling gates. A sequence of gates forms a quantum circuit, and measurement samples bitstrings from the output distribution. Quantum processors are noisy. Coherent errors arise from calibration imprecision and crosstalk, while stochastic errors from environmental coupling are characterized by $T_1$ and $T_2$ timescales. Both compound with circuit depth. These noise sources persist in fault-tolerant systems. Error correction reduces but does not eliminate the impact of qubit quality heterogeneity on logical error rates~\cite{fowler2012surface}. Mapping to higher-fidelity qubits remains desirable in the error-corrected regime.

\vspace{2mm}

\noindent\textbf{Variational algorithms and hybrid optimization.}
VQAs use a parameterized circuit (ansatz) with trainable rotation angles $\boldsymbol{\theta}$ to prepare states in a $2^n$-dimensional Hilbert space~\cite{Preskill2018, cerezo2021variational}. Execution follows a hybrid loop (Fig.~\ref{fig:background}). The ansatz runs on the quantum processor, measurements yield a cost $E(\boldsymbol{\theta}) = \langle \psi(\boldsymbol{\theta}) | H | \psi(\boldsymbol{\theta}) \rangle$ for a problem-specific Hamiltonian $H$, and a classical optimizer on HPC resources updates $\boldsymbol{\theta}$. Gradients are estimated via the parameter-shift rule~\cite{Mitarai2018, Schuld2019}. The quantum processor evaluates a cost over an exponentially large space that classical machines cannot efficiently represent, while the classical optimizer navigates through low-dimensional parameter updates. VQAs include VQE for quantum chemistry~\cite{Peruzzo2014}, QAOA for combinatorial optimization~\cite{Farhi2014}, and QNNs for quantum machine learning~\cite{Havlivcek2019}.

\vspace{2mm}

\noindent\textbf{Transpilation and physical mapping.}
Quantum processors have a fixed topology where each qubit interacts with few neighbors. Transpilation compiles a logical circuit into a physical circuit by selecting a layout, inserting SWAPs, decomposing into native gates, and optimizing. Physical qubits differ in error rates and coherence, so different layouts (Fig.~\ref{fig:background}) produce different noise exposure. We refer to a specific logical-to-physical assignment as a \emph{physical mapping} (or \emph{map}). The divergence between a map's output from an ideal noiseless simulation is \emph{total variation distance} (TVD). Transpilers select a map once by gate count or depth minimization~\cite{hua2023qasmtrans}.

In a variational optimization loop, the map is not a one-time decision. It affects the measured loss and gradients at every iteration. As parameters evolve and device noise fluctuates, the same map produces different noise exposure across iterations.

\begin{figure}[t] 
\centering
\includegraphics[scale = .3]{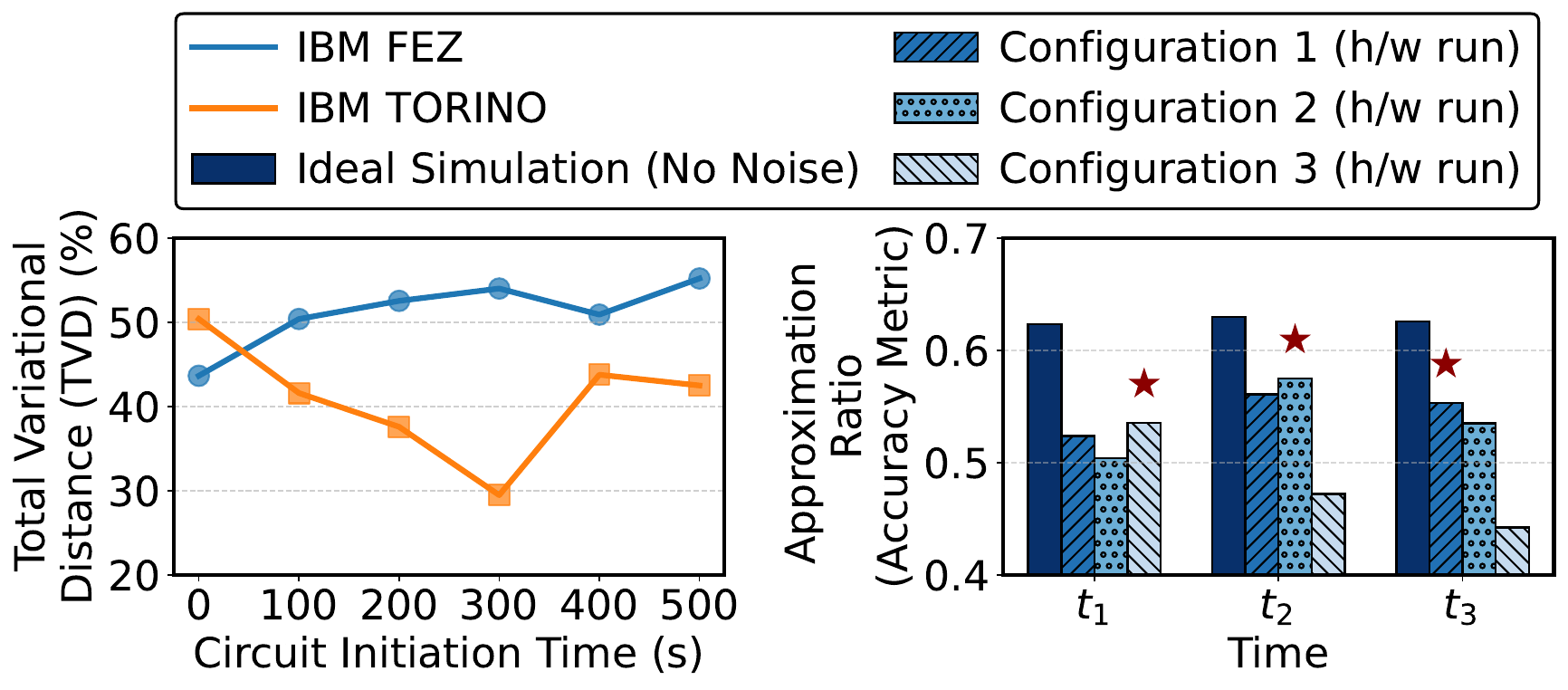}
\caption{\textit{TVD (lower is better) fluctuates when the same circuit is initiated at different times on quantum hardware.  Relative rankings of the performance metric for parameter configurations shift over time on real hardware. Stars mark the best configuration at each time instant on quantum hardware (h/w).}}
\label{fig:motivation_1}
\vspace{-6mm}
\end{figure}

To quantify how noise affects variational circuit optimization in practice, we run a QAOA MaxCut circuit on IBM Fez and IBM Torino and measure TVD against the ideal (noiseless) simulation output at multiple time points (experimental details in Sec.~\ref{sec:methodology}). The noiseless output comes from an \emph{ideal simulation}, a classical state-vector computation of the output a noise-free processor would produce, feasible only for circuits up to a certain qubit count before the $2^n$ state space exceeds classical memory~\cite{zhou2020limits}. QAOA encodes a combinatorial optimization objective into an ansatz whose rotation angles $\boldsymbol{\theta}$ are tuned to maximize the \emph{approximation ratio} (the performance metric for MaxCut, higher is better). A \emph{configuration} is a specific set of angle values. We select the three configurations with the highest approximation ratios under ideal simulation and track how their ranking changes on real hardware over time. The parameter-shift rule computes gradients by evaluating two shifted configurations and comparing their costs, so the optimizer makes correct updates only if the relative ranking of configurations on hardware matches the noiseless ranking.

Fig.~\ref{fig:motivation_1} (left) shows TVD measured by initiating the same circuit at multiple points within a time interval on both machines. Even over this short interval, TVD fluctuates due to qubit drift and thermal variation, and the user and transpiler have no control over this. On Fig.~\ref{fig:motivation_1} (right), we pick three time instants -- $t_1$, $t_2$, $t_3$ from this interval. The ideal simulation's approximation ratio for each configuration is stable across all three (solid dark bars). On real hardware (IBM Fez), the ranking shifts. Configuration~3 is best at $t_1$, Configuration~2 overtakes it at $t_2$, and the ordering changes again at $t_3$ (stars mark the best configuration at each time). The noise has changed, and with it the distortion on each configuration's output. Because these comparisons drive gradient computation, ranking errors perturb the update direction and accumulate across optimization iterations.

\begin{tcolorbox}
\textbf{Observation 1.} \textit{Hardware noise drifts over time, and this drift changes the relative performance ranking of parameter configurations on noisy hardware. Since variational optimizers rely on relative comparisons to compute gradients, noise-induced ranking inversions directly mislead parameter updates and destabilize optimization.}
\end{tcolorbox}

\begin{figure}[t] 
\centering
\includegraphics[scale = .3]{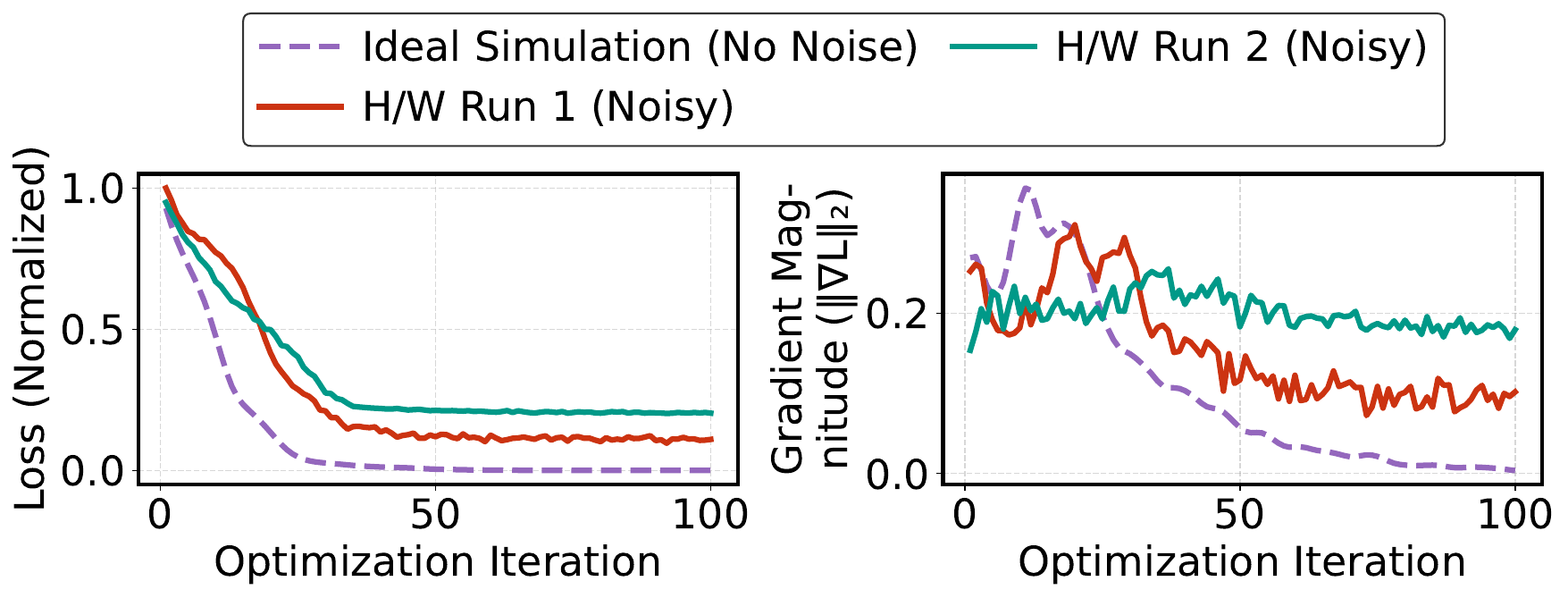}
\vspace{-4mm}
\caption{\textit{Ideal simulation loss converges to near zero and gradients decay. On noisy quantum hardware (h/w), loss plateaus and gradients remain elevated and erratic.}}
\label{fig:loss_grad_plot}
\vspace{-6mm}
\end{figure}

Ranking corruption has a direct consequence on the optimization trajectory. To show this, we optimize a QNN on IBM Fez and compare against an ideal simulation (Sec.~\ref{sec:methodology}). In the ideal simulation (Fig.~\ref{fig:loss_grad_plot}, left), loss drops to near zero. Gradient magnitude (Fig.~\ref{fig:loss_grad_plot}, right) decays as the optimizer nears the optimum value. While performing different QNN optimization runs on quantum hardware (Run 1 and Run 2), the loss for both runs plateaus well above the ideal minimum, and at different levels, because each run encounters a different noise trajectory as device conditions shift over time (Observation~1).  The gradient magnitude stays elevated and erratic because noise distorts the cost at every iteration, and the parameter-shift rule produces corrupted gradient estimates. These noisy gradients remain large near the true optimum, and the optimizer continues to update parameters without convergence. The loss plateaus because gradient noise exceeds the true signal. Other VQAs exhibit the same behavior.

\begin{tcolorbox}
\textbf{Observation 2.} \textit{Hardware noise distorts the measured loss and its gradients. The optimizer receives unreliable update signals throughout optimization, which prevents convergence to the true optimal solution.}
\end{tcolorbox}

\begin{figure}[t]
    \centering
    \begin{minipage}[t]{0.36\columnwidth}
        \vspace{0pt}
        \centering
        \scriptsize
        \begin{tabular}{cc}
            \toprule
            \makecell{Optimization \\ Iteration} & \makecell{Winning\\Map (in\\terms of TVD)} \\
            \midrule
            \cellcolor{itergray}1   & \cellcolor{mapone}Map 1 \\
            \cellcolor{itergray}20  & \cellcolor{maptwo}Map 2 \\
            \cellcolor{itergray}40  & \cellcolor{mapthree}Map 3 \\
            \cellcolor{itergray}60  & \cellcolor{mapthree}Map 3 \\
            \cellcolor{itergray}80  & \cellcolor{mapone}Map 1 \\
            \cellcolor{itergray}100 & \cellcolor{maptwo}Map 2 \\
            \bottomrule
        \end{tabular}
    \end{minipage}
    \hfill
    \begin{minipage}[t]{0.57\columnwidth}
        \vspace{0pt}
        \centering
        \includegraphics[width=\linewidth]{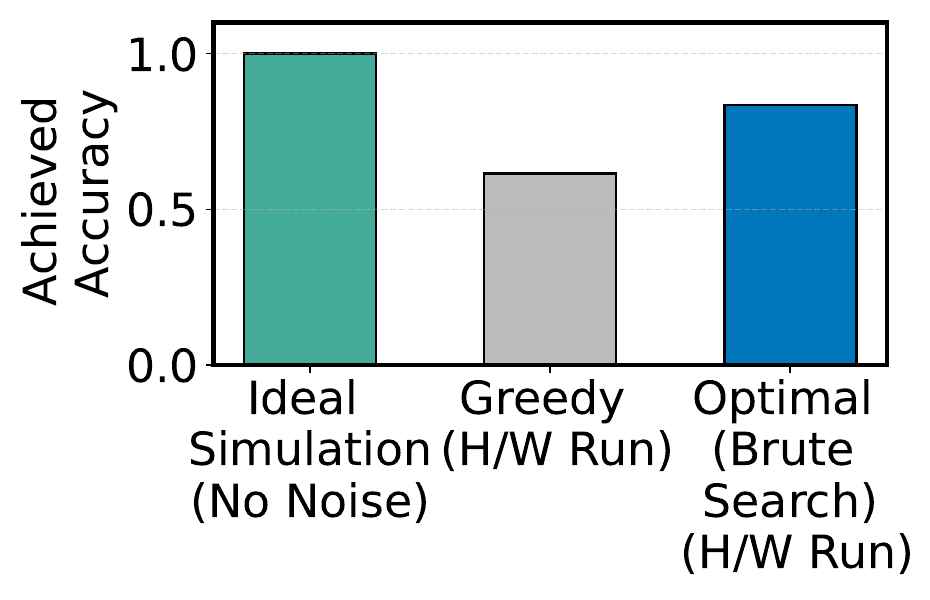}
    \end{minipage}
    \vspace{-1mm}
    \caption{\textit{TVD of transpiler-selected maps fluctuates across iterations, with the better map changing over time (left). Greedy map selection performs sub-optimally; the lowest per-iteration TVD does not lead to the best accuracy (right).}}
    \label{fig:accuracy_plot}
    \vspace{-6mm}
\end{figure}

Given that noise corrupts gradients and destabilizes optimization, an intuitive solution is to select the highest-fidelity map and fix it for the entire optimization. In practice, a circuit admits hundreds of valid physical maps due to different layout and routing choices on the device topology. Fig.~\ref{fig:accuracy_plot} (left) shows three representative maps selected from among the top candidates for a QNN on IBM Fez. The map with the lowest TVD changes across iterations. Device noise drifts between iterations and the parameters evolve~\cite{patel2020experimental}, which changes how the circuit interacts with each map's noise profile. No single map stays best. Greedy per-iteration selection (always pick the lowest-TVD map) is also sub-optimal (Fig.~\ref{fig:accuracy_plot}, right). Each iteration switches to a different map, so the noise profile the optimizer sees shifts between consecutive parameter updates. Updates computed under one noise profile are evaluated under a different one, and that mismatch compounds over iterations. An exhaustive brute-force search over all candidate maps at every iteration recovers much of the gap to the ideal, but the combinatorial cost of evaluating hundreds of maps per iteration makes this infeasible within realistic quantum budgets. The brute-force search implicitly selects maps with consistent noise exposure, while the greedy solution tries to improve per-iteration fidelity and sacrifices consistency.

What matters is not the lowest TVD at an isolated iteration, but whether TVD remains stable across iterations. Stable TVD keeps the noise distortion on the cost function consistent, which preserves configuration rankings (Observation~1) and gradient reliability (Observation~2). This stability needs to be achieved efficiently within quantum computing budgets.

\begin{tcolorbox}
\textbf{Observation 3.} \textit{The TVD of any physical map fluctuates across optimization iterations, and the best-performing map changes over time. Stabilizing TVD across iterations is necessary for reliable gradient computation in VQAs.}
\end{tcolorbox}

These observations motivate \sol{}, a control plane that stabilizes noise exposure across iterations and adapts map selection during both optimization and inference.

\section{\sol{}' Design}
\label{sec:design}

\noindent \sol{} operates as a control plane across three stages of VQA execution -- transpilation, circuit execution, and parameter update. During optimization, it selects physical maps per-iteration and gates gradient updates based on observed noise conditions on the quantum processor. At inference time, it re-computes map selection when hardware has drifted, or the backend has changed. We define the control abstraction below.

\subsection{Control Abstraction}
\label{sec:control_abstraction}
\noindent At each iteration $t$, \sol{} maintains candidate physical maps $C_t$, an estimated TVD $d_t(m)$ for each probed map $m \in C_t$, and a target TVD $\tau_t$ representing the current noise regime. \textit{Note that estimated TVD is derived from agreement among co-located outputs (Sec.~\ref{sec:consensus}), not from an ideal simulator.}

\vspace{2mm}

\noindent\textbf{Control objective: low and stable TVD.}
\sol{} selects maps with low TVD while maintaining stability across iterations. Low TVD alone is insufficient: Observation~3 shows that picking the lowest-TVD map each iteration leads to TVD swings and poor outcomes. The problem is the change between consecutive iterations, not the absolute value. The stability error of a candidate map is: $e_t(m) = |d_t(m) - \tau_t|$.

The target evolves as an exponential moving average:
\begin{equation}
\small \textstyle \tau_t = \alpha \, \tau_{t-1} + (1 - \alpha) \, d_{t-1}(m_{t-1})
\label{eq:target_tvd}
\end{equation}
$m_{t-1}$ is the map selected at the previous iteration. Among low-TVD candidates, the map with the lowest $e_t(m)$ is selected.

\vspace{2mm}

\noindent\textbf{Invariant: conditional gradient trust.}
Some iterations will still produce TVD that deviates from the target (e.g., a sudden calibration shift). The gradient from such an iteration reflects inconsistent noise conditions, and committing it can undo progress. \sol{} maintains a binary state $g_t \in \{0,1\}$ that determines whether the gradient bundle executes, and a continuous weight $\rho_t \in [0,1]$ that controls how strongly the gradient and estimated TVD are incorporated. The optimizer is unchanged. Gating and weighting are applied in the control plane. We formalize $g_t$ and $\rho_t$ in Sec.~\ref{sec:trust}. Together, low-and-stable TVD selection and conditional gradient trust are the properties that the design elements of \sol{} maintain.

\begin{figure}[t]
\centering
\includegraphics[width=\columnwidth] {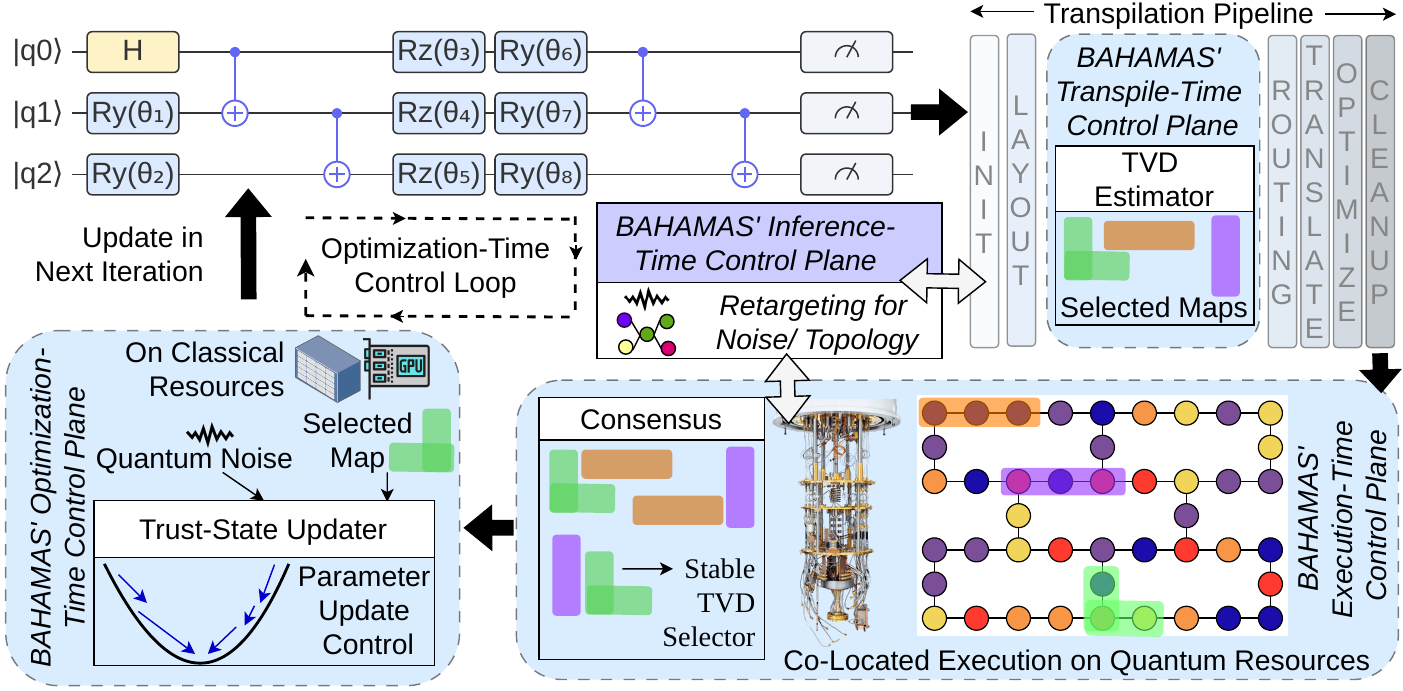}
\caption{\textit{\sol{}' control plane spans transpilation, execution, and parameter update. The TVD estimator shortlists $K$ low-TVD maps at transpile time. Co-located execution probes them under shared noise, and the stable TVD selector picks the best match to the target. The trust-state updater gates gradient updates on classical resources. In inference, map selection is retargeted for drifted noise or new topology.}}
\label{fig:overview}
\vspace{-4mm}
\end{figure}

\subsection{System Overview}
\label{sec:overview}

\noindent Fig.~\ref{fig:overview} shows \sol{}' architecture. The control plane spans three stages and operates in two modes: an optimization-time control loop and an inference-time retargeting mode.

\vspace{2mm}

\noindent\textbf{Optimization-time control loop.}
At transpile time, the transpiler generates candidate maps from the device topology. This set can be large (hundreds of valid maps for a given circuit), so a \emph{TVD estimator} within the transpile-time control plane ranks them using backend calibration data and consensus-estimated TVD from prior iterations, and outputs a shortlist of $K$ low-TVD candidates. These $K$ maps are then handed to the execution-time control plane, which uses \emph{co-located execution}: all $K$ maps of the same circuit (with the same parameters) are packed onto the chip in a single job, so their outputs reflect identical noise conditions. A \emph{consensus mechanism} estimates pairwise TVD among the $K$ output distributions. The map that best aligns with the others serves as a reference, and each map's TVD is estimated relative to it, eliminating the need for an ideal simulator. The \emph{stable TVD selector} then picks, among these low-TVD candidates, the map whose estimated TVD is closest to the target $\tau_t$ (Eq.~\ref{eq:target_tvd}). The selected map's output is passed to the optimization-time control plane, where the \emph{trust-state updater} checks whether the stability error is within tolerance and whether the co-located outputs sufficiently agree. If both hold, the gradient update proceeds as normal; otherwise, it is suppressed for that iteration. After each iteration, the estimated TVDs are fed back to the estimator. As the estimator accumulates observations, its ranking accuracy improves, and $K$ can be reduced over time.

\vspace{2mm}
\noindent\textbf{Inference-time retargeting.}
During optimization, \sol{} records the mean training-time estimated TVD and the coupling structure of maps selected during stable iterations. At inference time, the \emph{inference-time control plane} uses these to select a map without retraining. On the same backend after noise drift, it selects the map whose estimated TVD best matches the recorded training-time TVD. On a different backend with a different topology, it jointly matches both TVD and the coupling structure.

\subsection{\sol{}' Optimization-Time Control Plane}
\label{sec:opt_control}

\noindent The control abstraction in Sec.~\ref{sec:control_abstraction} requires estimated TVD values $d_t(m)$ at every iteration. The candidate space $\mathcal{C}_t$ can contain hundreds of valid maps, and probing all of them is not feasible. \sol{} first prunes $\mathcal{C}_t$ to a small probe set $P_t \subseteq \mathcal{C}_t$ with $K_t = |P_t| \ll |\mathcal{C}_t|$ using a classical TVD estimator, then executes only this shortlist on the quantum processor. The estimator ranks maps. The stable selector (Sec.~\ref{sec:stable_selector}) makes a final decision based on the actual estimated TVD. The estimator does not directly optimize TVD. It primarily makes sure that maps whose TVD is likely to be near the target (Eq.~\ref{eq:target_tvd}) are retained for probing.

\vspace{2mm}

\subsubsection{Candidate Map Space}
\label{sec:candidate_space}

Different layout and routing choices produce many valid physical maps of the same ansatz on a given backend. A map $m$ is characterized by its \emph{routed hardware footprint}: $w_q(m)$ is the normalized exposure to physical qubit $q$, and $w_e(m)$ is the normalized exposure to coupler $e$. These capture how the circuit's gates, SWAPs, and measurements are distributed across the device. Because the ansatz is fixed across iterations, footprints are computed once.

\begin{figure}[t]
\centering
\includegraphics[width=\columnwidth] {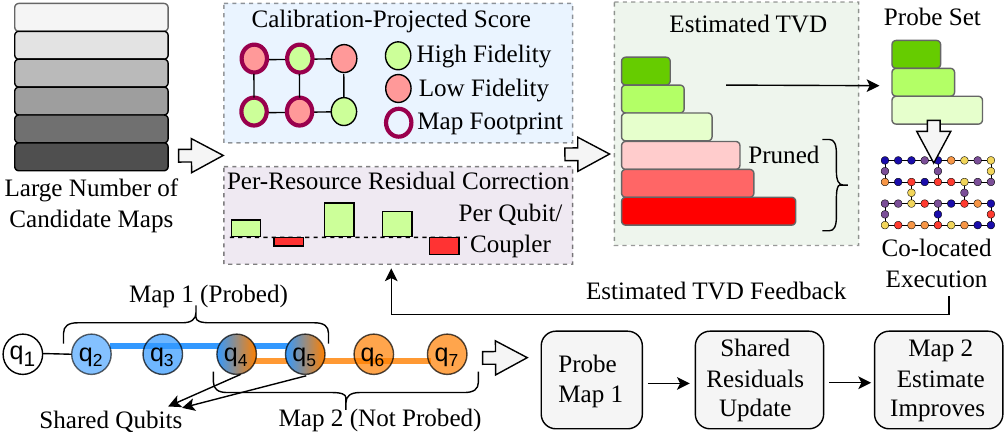}
\vspace{-4mm}
\caption{\textit{TVD estimator combines device calibration data with per-resource residuals to rank and prune candidate maps to a small probe set. Residuals are updated each iteration from estimated TVDs. Because residuals are attached to physical qubits and couplers, probing one map improves estimates for all maps that share physical resources with it.}}
\label{fig:tvd_estimator}
\vspace{-6mm}
\end{figure}

\vspace{2mm}

\subsubsection{TVD Estimator}
\label{sec:estimator}

At iteration $t$, the backend provides calibration data with per-qubit risk $x_t^q(q)$ and per-coupler risk $x_t^e(e)$. The calibration-projected exposure score for map $m$ is:
\begin{equation}
\small \textstyle s_t(m) = \sum_q w_q(m) \, x_t^q(q) + \sum_e w_e(m) \, x_t^e(e)
\label{eq:calibration_projected}
\end{equation}
This score weights each map's hardware footprint by the device's current error rates (Fig.~\ref{fig:tvd_estimator}). It is cheap to compute but incomplete, as calibration data does not fully capture circuit-level estimated TVD, and the mismatch drifts as device conditions change. An alternative is to train a neural network or gradient-boosted model offline on historical (map, TVD) pairs. In our experiments, such models rank well on the backend they were trained on, but degrade within hours as noise drifts, and they need to be retrained for each new backend. A simpler online approach better fits this setting. The estimator maintains per-qubit residuals $u_t(q)$, per-coupler residuals $v_t(e)$, and global terms $a_t, b_t$:
\begin{equation}
\small \textstyle \hat{d}_t(m) = a_t + b_t \, s_t(m) + \sum_q w_q(m) \, u_t(q) + \sum_e w_e(m) \, v_t(e)
\label{eq:estimated_tvd}
\end{equation}
The estimator does not need an accurate absolute TVD. The stable selector (Sec.~\ref{sec:stable_selector}) only needs the relative ranking of maps to be correct. Ranking requires that estimation errors are roughly consistent across maps so that ordering is preserved even if values are off. Global terms $a_t, b_t$ absorb systematic bias shared across all maps, which preserves relative ordering. Residuals are attached to physical resources, not maps, so a measurement from one probed map improves ranking for all maps sharing the same qubits and couplers (Fig.~\ref{fig:tvd_estimator}). A per-map residual design is an alternative, but unprobed maps receive no updates and drift in ranking as noise changes. Per-resource residuals avoid this because any map touching qubit $q$ contributes to $u_t(q)$.

After co-located execution returns $d_t(m)$ for $P_t$, estimation errors update the residuals for touched resources via exponential weighting proportional to estimation error and resource exposure. Note that this process requires no offline training. The probe budget $K_t$ adapts to the TVD estimator's uncertainty. An error envelope $\delta_t$ tracks what has been the recent estimation accuracy. Let $m^{(1)} = \arg\min_{m \in \mathcal{C}_t} \hat{d}_t(m)$. Probe set includes maps within $\delta_t$ of $m^{(1)}$:

{\small
\begin{align*}
\textstyle P_t &= \{ m \in \mathcal{C}_t \mid \hat{d}_t(m) \leq \hat{d}_t(m^{(1)}) + \delta_t \} \\
\textstyle K_t &= \mathrm{clip}(|P_t|,\; K_{\min},\; K_{\max})
\end{align*}}


This allows execution on a small set of candidates while preserving the map that satisfies the stability objective with high probability. Early in optimization, $\delta_t$ is larger. As residuals improve, it shrinks, and overhead decreases.

\textcolor{black}{The estimator's task is correct \emph{ranking}, not absolute TVD, and it is robust in this respect. From our experiments (discussed in Sec.~\ref{sec:methodology}), we observe that as per-resource residuals accumulate, the rank correlation between estimated and measured map TVD rises on average from 0.70 to 0.92 (Spearman), and the selected map falls within the lowest-TVD probe set in 92\% of iterations. After an injected calibration shift, ranking recovers within a few iterations. An offline predictor trained on historical (map, TVD) pairs starts near 0.90 but degrades to about 0.60 under drift and must be retrained per backend, whereas the online estimator holds near 0.90 without retraining. This is why \sol{} requires no simulator and no offline training.}

\vspace{2mm}

\subsubsection{Co-Located Execution}
\label{sec:colocated}
The estimator produces a probe set $P_t$ of $K_t$ maps. These now need to be evaluated on the quantum processor to obtain TVD estimations via consensus (Sec.~\ref{sec:consensus}). Running them sequentially would expose each map to a different noise snapshot, thereby conflating map quality with temporal drift. \sol{} avoids this by packing all $K_t$ maps onto the chip in a single job. Each map in $P_t$ already occupies a different subset of physical qubits by construction (that is what distinguishes one map from another). The logical circuit is small relative to the full device, so multiple transpiled copies fit on non-overlapping qubit subsets simultaneously. The $K_t$ copies execute under the same noise conditions, so any difference in their outputs is due to the map alone. This also amortizes probing cost to a single job submission and execution window for all $K_t$ candidates.

\vspace{2mm}

\subsubsection{Consensus-Based Reference}
\label{sec:consensus}
Co-located execution produces $K_t$ output distributions $\{P_1, \ldots, P_{K_t}\}$. Computing each map's TVD requires a reference distribution. For small circuits, a noiseless simulator can provide this, but simulation cost grows exponentially with qubit count and becomes infeasible as circuits scale. \sol{} avoids this dependency entirely by constructing a reference from the co-located outputs themselves. The consensus mechanism identifies the most representative output:
\begin{equation}
\small \textstyle \hat{c}_t = \arg\min_{i \in [K_t]} \sum_{j \neq i} \mathrm{TVD}(P_i, P_j)
\label{eq:consensus}
\end{equation}
Each map's estimated TVD is then $d_t(m_i) = \mathrm{TVD}(P_i, P_{\hat{c}_t})$. All $K_t$ maps execute the same logical circuit with the same parameters, so they all attempt to produce the same ideal output. The only difference is the noise each map experiences. The map with the least distortion produces an output closest to what every map is trying to compute and therefore tends to have the smallest total disagreement with the group. The consensus reference is thus biased toward the least noisy output, and the distance from it correlates with noise severity across maps. This does not recover absolute TVD relative to the true noiseless output, but it preserves the relative ordering of maps by noise, which is what the stable selector needs. The degree of agreement also provides a confidence signal: \sol{} defines consensus confidence $\gamma_t$ as the inverse of the mean pairwise TVD among the outputs. A high $\gamma_t$ indicates that the outputs were consistent and that the TVD estimations are reliable. Low $\gamma_t$ flags an unusual noise event. Both $d_t(m)$ and $\gamma_t$ feed into the next two stages.

\vspace{2mm}

\subsubsection{Stable TVD Selector}
\label{sec:stable_selector}
Given estimated TVDs from the consensus step, the selector picks the map for this iteration's gradient computation:
\begin{equation}
\textstyle m_t = \arg\min_{m \in P_t} |d_t(m) - \tau_t|
\label{eq:stable_select}
\end{equation}
The target $\tau_t$ (Eq.~\ref{eq:target_tvd}) tracks the recent operating point, and the selector finds the map that best preserves it. If multiple maps in $P_t$ are close to the target, the one with the lower absolute TVD is preferred. The selected map $m_t$ is locked for the full gradient bundle at this iteration, including across the paired parameter-shift evaluations for each partial derivative, because switching maps mid-gradient would mix noise from different physical configurations. After selection, \sol{} computes the stability error $e_t(m_t)$. The trust-state updater then decides whether to execute the gradient bundle, how strongly the resulting gradient should influence the optimizer, and how much the TVD should influence the next target.

\begin{figure}[t]
\centering
\includegraphics[width=\columnwidth] {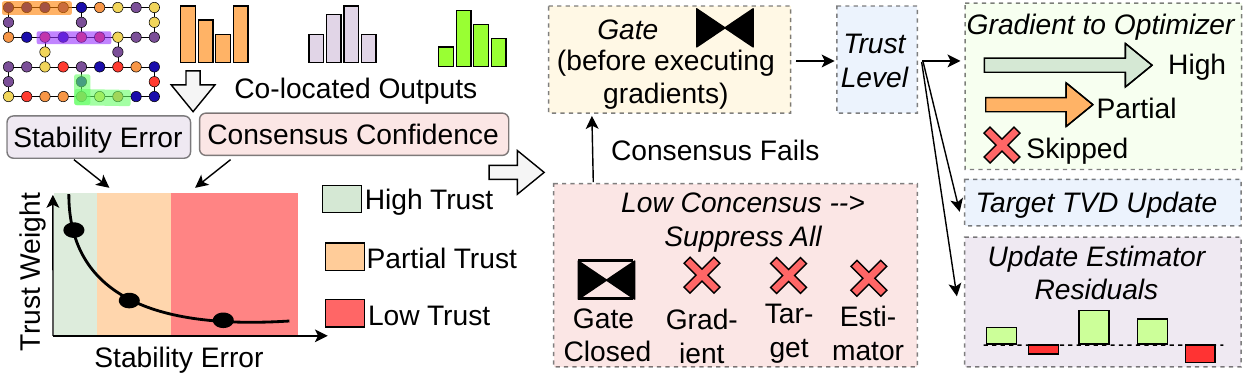}
\vspace{-4mm}
\caption{\textit{Trust-aware gradient control. Co-located outputs determine consensus confidence and stability error, which jointly set the trust weight via exponential decay. Trust gates and scales gradient updates to the optimizer, controls how quickly the target TVD adapts, and governs estimator residual updates. Low consensus closes the gate and suppresses all updates.}}
\label{fig:trust_aware}
\vspace{-6mm}
\end{figure}

\vspace{2mm}

\subsubsection{Trust-State Updater}
\label{sec:trust}

Stable map selection reduces noise swings, but not every iteration should spend gradient budget or contribute equally to optimization. \sol{} computes trust from two quantities already produced by the control loop: the stability error $e_t(m_t)$ and the consensus confidence $\gamma_t$ (Fig.~\ref{fig:trust_aware}, left). Trust is evaluated before launching the gradient bundle (the gate in Fig.~\ref{fig:trust_aware}), so unnecessary quantum executions under unreliable conditions are avoided. Trust is computed once per iteration and applied to the entire gradient bundle, because all partial derivatives share the same map and noise realization.

Consensus confidence acts as a hard gate: $c_t = \mathbf{1}[\gamma_t \ge \gamma_{\min}]$. If the co-located outputs disagree, estimated TVDs are unreliable, and the gate closes, suppressing all downstream updates. Conditioned on a reliable consensus, the trust weight decays smoothly with the stability error:
\begin{equation}
\small \textstyle \rho_t = c_t \cdot \exp\!\left(-\frac{e_t(m_t)}{\epsilon}\right), \quad g_t = \mathbf{1}[\gamma_t \ge \gamma_{\min} \;\land\; e_t(m_t) \le \epsilon]
\label{eq:trust}
\end{equation}
As shown in the trust weight curve (Fig.~\ref{fig:trust_aware}), small stability error yields high trust, moderate error yields partial trust, and large error drives trust toward zero. The exponential form preserves high trust near the target while rapidly penalizing larger deviations. Alternatively, we evaluated linear decay ($\rho_t = c_t \cdot \max(0, 1 - e_t/\epsilon)$) and a hard threshold ($\rho_t = g_t$). Linear decay assigns near-zero weights to steps close to the admissible boundary, thereby wasting usable gradient information. The hard threshold discards all partial trust and treats a step at $e_t = 0.9\epsilon$ identically to one at $e_t = 0$. We observed that the exponential form avoids both extremes and produced the best optimization outcomes in our experiments.

The trust level controls three downstream paths (Fig.~\ref{fig:trust_aware}, right). First, the gradient to the optimizer: if $g_t = 0$, \sol{} skips the parameter-shift execution entirely; if $g_t = 1$, the full gradient bundle runs on $m_t$ and the optimizer receives $\tilde{\nabla}_t = \rho_t \nabla_t$, scaled from full strength (high trust) down to partial strength or skipped depending on the trust level. The optimizer itself is unchanged. A naive alternative is to commit every gradient at full weight regardless of noise conditions, equivalent to always setting $\rho_t = 1$. We evaluate this as the \textit{always trusting gradients} baseline in Sec.~\ref{sec:eval}. Second, the target TVD update: \sol{} replaces the update in Eq.~\ref{eq:target_tvd} with $\tau_{t+1} = \alpha \tau_t + (1 - \alpha)[\rho_t d_t(m_t) + (1 - \rho_t)\tau_t]$. When trust is high, the target adapts to the new observation. When trust is low, the target moves only gradually. When consensus fails, the target is frozen. Third, the estimator residuals are updated whenever $\gamma_t \ge \gamma_{\min}$, even if $g_t = 0$, because a high-consensus probe set with large stability error still reveals which resources have drifted. Low consensus suppresses all three paths: gradient, target, and estimator updates are all suppressed, and \sol{} widens the next probe set to recover a reliable operating point.

\vspace{2mm}

\subsubsection{Online Feedback and Budget Adaptation}
\label{sec:feedback}

After each iteration, estimated TVDs from the co-located execution feed back to the TVD estimator. Per-resource residuals are updated for all qubits and couplers touched by the probed maps. As residuals accumulate over iterations, the estimator's ranking accuracy improves, and the error envelope $\delta_t$ shrinks. The probe set $P_t$ contracts accordingly. Early iterations probe broadly because the estimator has little history, and later iterations probe fewer maps as the estimator learns the backend. The per-iteration overhead of the optimization-time control plane, therefore, decreases over the course of optimization. No offline training or simulator access is required at any point.

The optimization-time control loop also produces a byproduct that \sol{} reuses at inference. These are the mean target TVD and the coupling structure of maps selected during stable iterations. Next, we describe how \sol{} uses these for inference-time retargeting.

\begin{figure}[t]
\centering
\includegraphics[width=\columnwidth]{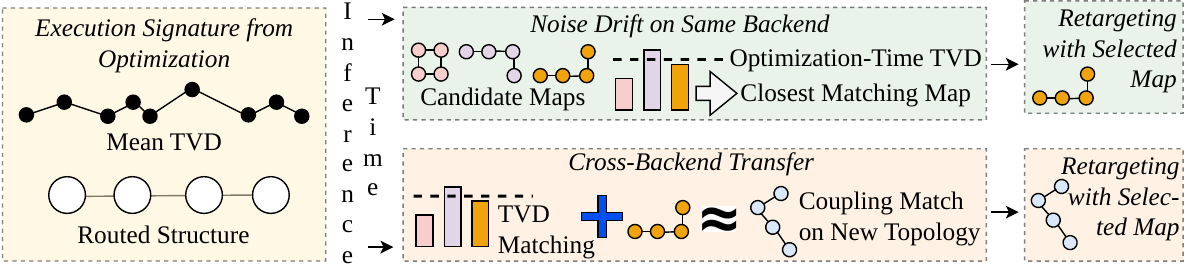}
\vspace{-4mm}
\caption{\textit{The execution signature from optimization drives map selection during inference. Same-backend drift matches candidate maps to the TVD used at optimization time. Cross-backend transfer jointly matches TVD and coupling structure.}}
\label{fig:inference_retargeting}
\vspace{-6mm}
\end{figure}

\subsection{\sol{}' Control Plane for Inference-Time Retargeting}
\label{sec:inference}

\noindent The control abstraction used during optimization can be reused during inference/deployment to preserve the optimization-time execution signature without reoptimization. During optimization, \sol{} constructs an execution signature from trusted iterations ($g_t = 1$), which includes the mean estimated TVD $\bar{\tau}_{\mathrm{opt}}$ and a normalized routed interaction vector $r_{\mathrm{opt}}$ that records the path length each two-qubit interaction required under the selected maps. This signature captures both the noise level and the stable operating point the optimizer converged under, so that inference-time map selection preserves the same stability the control plane maintained during optimization.

\vspace{2mm}

\subsubsection{Same-Backend Drift}
\label{sec:inf_same}

When deployed on the same backend after noise drift, \sol{} refreshes the calibration snapshot, re-scores the candidate set $\mathcal{C}_{\mathrm{inf}}$, and selects the map whose estimated TVD best matches the optimization-time operating point (Fig.~\ref{fig:inference_retargeting}, top):
\begin{equation}
\small \textstyle m^\star = \arg\min_{m \in \mathcal{C}_{\mathrm{inf}}} \left| \hat{d}_{\mathrm{inf}}(m) - \bar{\tau}_{\mathrm{opt}} \right|
\label{eq:inf_same}
\end{equation}
No reoptimization is needed. The selected map restores noise close to that under which the parameters were optimized.

\vspace{2mm}

\subsubsection{Cross-Backend Transfer}
\label{sec:inf_cross}

On a different backend, per-resource residuals do not transfer. \sol{} reuses only $(\bar{\tau}_{\mathrm{opt}}, r_{\mathrm{opt}})$. For each candidate map $m$ on the new backend, it computes a fresh TVD estimate $\hat{d}_{\mathrm{new}}(m)$ and a coupling distance $\Delta_{\mathrm{cpl}}(m, r_{\mathrm{opt}})$ (the $\ell_1$ distance between routed interaction vectors). As shown in Fig.~\ref{fig:inference_retargeting} (bottom), the selected map should preserve both the optimization-time noise level and the routed structure on the new topology:
\begin{equation}
\small \textstyle m^\star = \min_{m \in \mathcal{C}'} \lambda \left| \hat{d}_{\mathrm{new}}(m) - \bar{\tau}_{\mathrm{opt}} \right| + (1-\lambda)\,\Delta_{\mathrm{cpl}}(m, r_{\mathrm{opt}})
\label{eq:inf_cross}
\end{equation}
where $\mathcal{C}'$ is the candidate map set on the new backend. The first term preserves the noise level that the parameters expect. The second preserves the interaction structure used to optimize the parameters. TVD matching alone is insufficient because the same TVD can arise from different noise distributions over the circuit, and the parameters are sensitive to both. Because $r_{\mathrm{opt}}$ is indexed by logical interactions, the comparison is well-defined across topologies. Next, we discuss several steps taken to make  \sol{}' design robust at handling various corner cases that arise in practice.

\subsection{System Integration and Robustness}
\label{sec:robustness}

\noindent \sol{} wraps an existing transpiler and requires no changes to circuit syntax, optimizer, or parameter-shift code. When a signal is unreliable, \sol{} suppresses the decisions that depend on it and increases probing on next iteration.

\vspace{2mm}

\noindent\textbf{Weak consensus and insufficient shots.}
If co-located outputs disagree beyond the $\gamma_{\min}$ threshold, the consensus reference is unreliable and all downstream updates (gradient, target, estimator) are suppressed for that iteration. Low shot budgets produce the same effect by increasing variance in the output distributions, which \sol{} detects through $\gamma_t$. In both cases, \sol{} widens the probe set on the next iteration to increase the chance of restoring a stable operating point.

\vspace{2mm}

\noindent\textbf{Calibration staleness and backend failures.}
The TVD estimator depends on calibration data that can go stale between iterations. If the snapshot age exceeds a configurable threshold, \sol{} re-fetches calibration before reusing the last trusted map. If probing cannot complete due to transient backend errors or queue delays, \sol{} reuses the last trusted map and defers estimator updates to the next successful iteration. In both cases, the optimization loop continues without stalling. If co-location is unavailable, \sol{} probes sequentially within the smallest available time window and increases $K_t$ to compensate.

\vspace{2mm}

\noindent\textbf{Topology mismatch at inference.}
Cross-backend retargeting (Sec.~\ref{sec:inf_cross}) relies on coupling distance $\Delta_{\mathrm{cpl}}$. If coupling information is unavailable on the target backend or the topology difference is too large for this distance to be meaningful, \sol{} falls back to TVD-only matching (Eq.~\ref{eq:inf_same}), which requires only calibration data.

\subsection{\sol{}: Putting It All Together}
\label{sec:together}

\noindent Algorithm~\ref{alg:bahamas} summarizes the full \sol{} control plane. At each optimization iteration, the TVD estimator scores all candidate maps using calibration data and per-resource residuals, and prunes to a probe set of $K_t$ maps. These are co-located on the quantum processor in a single job. The consensus mechanism identifies the most representative output and measures each map's TVD relative to it. The stable selector picks the map closest to the target TVD. The trust-state updater then decides whether the gradient bundle should execute and at what strength, based on the stability error and consensus confidence. If trusted, the gradient is scaled and the target adapts. If not, the iteration is suppressed and the probe set is widened. After execution, measured TVDs feed back to the estimator to improve its accuracy and shrink the probe budget over time. At inference-time, \sol{} reuses the execution signature accumulated during optimization to retarget map selection on drifted or new hardware without reoptimization. Next, we discuss \sol{}' implementation.

\begin{algorithm}[t]
\caption{\sol{}' Control Plane}
\label{alg:bahamas}
{\scriptsize
\begin{algorithmic}[1]
\REQUIRE Ansatz, optimizer, backend, $\alpha, \epsilon, \gamma_{\min}$
\STATE Compute footprints $w_q(m), w_e(m)$ $\forall$ maps; init $\tau_0, (a_0, b_0, u_0, v_0)$
\STATE \textbf{for} each optimization iteration $t$ \textbf{do}
\STATE \quad Fetch calibration $x_t$; score $\hat{d}_t(m)$ via estimator (Eq.~\ref{eq:estimated_tvd}); form probe set $P_t$
\STATE \quad Co-locate $P_t$ on chip in one job; compute consensus $\hat{c}_t$ (Eq.~\ref{eq:consensus})
\STATE \quad Estimate $d_t(m)$ $\forall$ $m \in P_t$ relative to $\hat{c}_t$; compute confidence $\gamma_t$
\STATE \quad Select $m_t = \arg\min_{m \in P_t} |d_t(m) - \tau_t|$; compute $e_t(m_t), \rho_t, g_t$ (Eq.~\ref{eq:trust})
\STATE \quad \textbf{if} $g_t = 1$: execute gradient on $m_t$, pass $\rho_t \nabla_t$ to optimizer, update $\tau_{t+1}$
\STATE \quad Update estimator residuals for resources in $P_t$; record signature if $g_t = 1$
\STATE \textbf{Inference:} retarget using $(\bar{\tau}_{\mathrm{opt}}, r_{\mathrm{opt}})$ via Eq.~\ref{eq:inf_same} or \ref{eq:inf_cross}
\end{algorithmic}}
\end{algorithm}

\section{Implementation}
\label{sec:implementation}

\noindent\textbf{Control-plane runtime.}
\sol{} is implemented as a library between a frontend quantum SDK and a backend execution API. A persistent state store maintains the estimator state $(\tau_t, \delta_t, a_t, b_t, u_t, v_t)$ and execution signature $(\bar{\tau}_{\mathrm{opt}}, r_{\mathrm{opt}})$, scaling as $O(|Q| + |E|)$ independent of shots or optimization history. Transient trust values $(g_t, \rho_t)$ are computed per iteration and discarded. Optimizer and user circuit are unchanged.

\vspace{2mm}

\noindent\textbf{Cached artifacts and overhead.} The frontend adapter lowers the ansatz once and caches each map's routed template, footprint $(w_q, w_e)$, interaction vector $r(m)$, and stable \texttt{map\_id}. Subsequent iterations bind parameters without re-transpiling. Classical scoring and consensus complete in $<$55\,ms CPU time per iteration. The dominant quantum cost is one co-located probe batch ($\sim$10\% of per-iteration shots) plus one gradient bundle (only when $g_t = 1$). $K_t$ shrinks as the estimator improves, and skipped iterations recover gradient shots, yielding minimal shot overhead. Inference retargeting requires no quantum execution.

\vspace{2mm}

\noindent\textbf{User API and framework adapters.}
The control plane abstracts all internal decisions behind a narrow interface. The user provides the ansatz, initial parameters, a loss function, and an optimizer. \sol{} handles map selection, probing, consensus, trust gating, and feedback internally. A frontend adapter implements \texttt{to\_internal\_repr} and \texttt{generate\_candidates}, a backend adapter implements \texttt{get\_calibration}, \texttt{submit\_batch}, and \texttt{get\_counts}. Any SDK that can represent a parameterized circuit and submit shots can plug in. \sol{} ships adapters for Qiskit, PennyLane, and Cirq, with execution signatures indexed by logical interactions for portability across others.

\begin{lstlisting}[style=bahamas]
# Direct or framework-wrapped initialization
ctrl = BAHAMAS(backend, min_probe_budget,
  max_probe_budget, target_smoothing_rate,
  stability_tolerance, consensus_threshold, 
  coupling_weight)
ctrl = BAHAMAS.from_framework(backend)
# Optimization and inference
params = ctrl.optimize(ansatz, initial_params,
  optimizer, loss_function)
result = ctrl.infer(ansatz, params)
\end{lstlisting}

\section{Experimental Methodology}
\label{sec:methodology}

\noindent\textbf{Hardware and execution.}
We evaluate \sol{} on different currently available IBM quantum processors on IBM Quantum Cloud, spanning three processor generations and two coupling topologies. IBM Fez and IBM Kingston are Heron r2 processors (156 qubits, heavy-hex lattice). IBM Pittsburgh is a Heron r3 processor (156 qubits, heavy-hex lattice), IBM's highest-performing Heron revision with improved coherence and gate fidelity. IBM Miami is a Nighthawk r1 processor (120 qubits, square lattice), IBM's early fault-tolerant architecture with higher connectivity and lower error rates than previous generations. All primary experiments run on real quantum hardware with 8,192 shots per circuit execution, sufficient for reliable cost function and gradient estimation across our benchmark sizes.  Each configuration (benchmark, technique, backend) is repeated 100 times over a 30-day period, with runs distributed across different times of day and calibration cycles to capture temporal noise variation. We report means and standard deviations over these runs. Higher repetition counts are cost-prohibitive on IBM Quantum Cloud due to per-shot billing and multiple iterations of VQAs. We verified through noisy simulation that increasing to 200 runs does not meaningfully reduce variance in the reported metrics. Two experiments that require controlled noise sweeps (Figs.~\ref{fig:noise_backends} (right) and~\ref{fig:noise_scaling_combined} (right)) use noisy simulation with calibration data imported from live backends and per-gate error rates scaled by a controlled factor, following standard methodology in the quantum systems literature~\cite{huo2025anchor, huo2025revisiting, ravi2021quantum}. These run on AMD Genoa 96-core server and NVIDIA H100NVL GPU.

\vspace{2mm}

\noindent\textbf{Benchmarks and VQA pipeline.}
We evaluate three VQA families across eight benchmarks. VQE targets ground-state energy on H$_2$ (4 qubits), HeH$^+$ (4 qubits), and H$_3^+$ (6 qubits) using EfficientSU2 with 3 repetitions~\cite{arrazola2021differentiable, ravi2022vaqem}. QAOA solves MaxCut on Cycle, Ladder, and Star graphs (one-layer ansatz, 6 qubits). QNN trains classifiers on MNIST (3 vs.\ 4) and FashionMNIST (Bag vs.\ Sandal) with 10 qubits and 30 parameters per layer. We scaled each benchmark up to 16 qubits; beyond that, TVD on current hardware rises to levels where outcomes are not representative regardless of map selection. \textcolor{black}{This 16-qubit ceiling is a property of current NISQ hardware and of the classical noiseless reference used for evaluation, not of \sol{}: beyond this scale the state-vector reference becomes classically intractable and device outputs stop being representative, while \sol{} itself uses no simulation and its cost grows only as $O(\text{qubits}+\text{couplers})$ (Sec.~\ref{sec:estimator}).}  This suite follows established VQA benchmarking practice~\cite{huo2025three, quetschlich2023mqt, lorenz2025systematic, blekos2024review}. All techniques use identical ansatz configurations and Qiskit transpiler optimization level 3 (highest level). Each benchmark follows standard hybrid loop: circuits constructed in Qiskit (v1.2.4)~\cite{javadi2024quantum}, executed on target backend, costs estimated from measurement counts and passed to COBYLA~\cite{powell2007view} via SciPy~\cite{virtanen2020scipy}. We verified with Adam and SGD (parameter-shift gradients) that relative improvement over baselines remains consistent, as \sol{} operates below the optimizer and does not modify the update rule.

\vspace{2mm}

\begin{figure}[t!]
\vspace{-4mm}
\centering
\includegraphics[scale=.29]{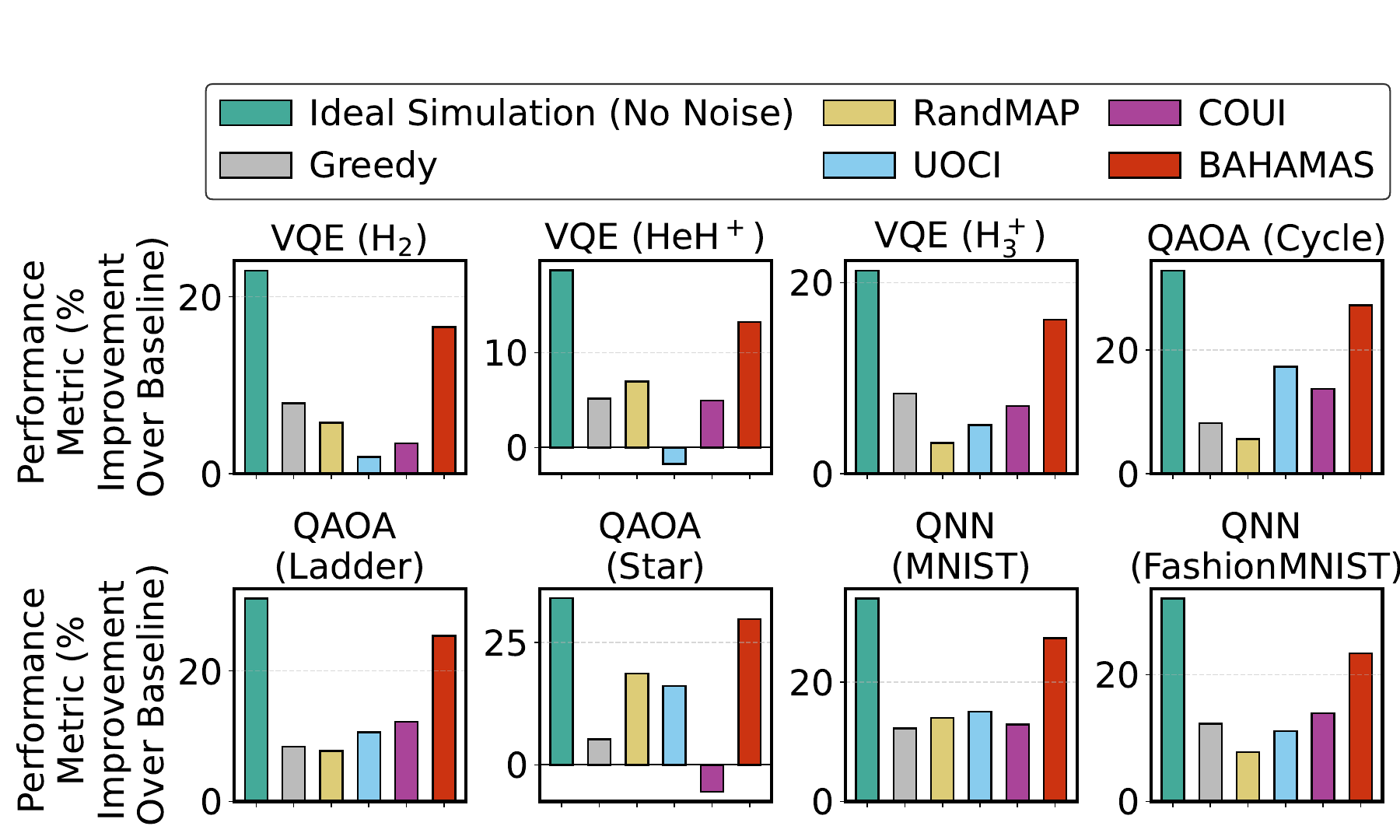}
\vspace{-4mm}
\caption{\textit{\sol{} improves on the performance metric (defined in Sec.~\ref{sec:methodology}), approaching the noiseless simulation bound.}}
\label{fig:performance_improvement}
\vspace{-6mm}
\end{figure}

\noindent\textbf{Competing solutions.} \textbf{Ideal Simulation (No Noise)} executes on a noiseless simulator (running on classical hardware) and serves as the \textit{Oracle upper bound}. All other configurations run on real quantum hardware. \textbf{Uncontrolled (Baseline)} \textit{represents how VQAs run today} -- the transpiler selects a single physical map before optimization begins, and that map is used for every iteration through completion with no noise-aware adaptation. This is the primary baseline. The remaining configurations are not independent techniques but partial configurations of \sol{} that enable or disable specific controls to isolate their individual contribution. \textbf{Greedy} enables \sol{}' per-iteration map selection but disables stability tracking and trust gating, always picking the lowest-TVD map. \textbf{RandMAP} replaces \sol{}' estimator and selector with random map choice at each iteration. \textbf{UOCI} (Uncontrolled Optimization, Controlled Inference) runs optimization without any control plane but enables \sol{}' inference-time retargeting. This isolates the value of optimization-time control. \textbf{COUI} (Controlled Optimization, Uncontrolled Inference) enables optimization-time control but reverts to default transpiler mapping at inference to isolate the value of retargeting. We additionally compare \sol{} against a prior noise-adaptive execution technique for VQAs, \textbf{QISMET}~\cite{ravi2023navigating}, which fixes a single physical map and reruns prior iterations to detect and reject parameter updates taken under drifted noise.

\vspace{2mm}

\noindent\textbf{Complementary technique.} No prior work directly addresses per-iteration noise-aware map selection with gradient trust for VQAs. The closest is \textbf{NEST}~\cite{huo2025three}, which targets a different problem -- improving VQA throughput by varying qubit fidelity across iterations and co-locating multiple jobs on the same backend. It operates at the job scheduling and qubit allocation level. \sol{} performs per-iteration map selection and gradient trust within a given allocation. The two techniques are complementary. NEST decides which chip region a job occupies, \sol{} decides which map to use within that region, and whether to trust the gradient or not. We evaluate them jointly to assess whether the benefits reinforce each other. Additionally, another solution, \textbf{DISQ}~\cite{zhang2023disq}, keeps a fixed map and skips iterations whose measured state is deemed unreliable. It reacts to noise on a fixed map, instead of controlling per-iteration map selection. This is why \sol{} is complementary to the DISQ technique. We evaluate the \sol{}+DISQ combination in Sec.~\ref{sec:eval}.

\vspace{2mm}

\noindent\textbf{Evaluation metrics.}
The primary metric (performance metric) is improvement over the uncontrolled baseline: Energy Gap for VQE, Approximation Ratio for QAOA/MaxCut, and classification accuracy for QNN (on a held-out test set not seen during optimization). Each configuration runs 100 independent optimization-inference pairs over a 30-day period, with runs interleaved across different calibration cycles and noise conditions. For each pair, optimization completes first and inference is performed days later so that device noise has measurably drifted. Parameters are frozen at inference and evaluation covers both the same backend (capturing temporal drift) and a different backend (capturing topology transfer). This separates optimization-time control from inference-time robustness under changed hardware conditions. We report standard deviation of the performance metric across these 100 pairs. Each pair encounters different noise conditions, so standard deviation directly measures sensitivity to temporal variation -- lower variance is a design goal of \sol{}.

\section{Evaluation}
\label{sec:eval}

\subsection{\sol{}' Performance Effectiveness}
\label{sec:eval_perf}

We first evaluate end-to-end performance across all eight benchmarks on IBM Pittsburgh, averaged over 100 optimization-inference pairs run over 30 days. \sol{} achieves an average of 22.39\% improvement over the baseline (no control during optimization and inference, standard way of executing VQAs) across all benchmarks (Fig.~\ref{fig:performance_improvement}), approaching the ideal noiseless simulation bounds. Greedy falls short because it chases per-iteration TVD without stability tracking (Observation~3). RandMAP performs worse, which confirms that map quality matters even without stability. The partial configurations confirm that optimization-time control is the dominant factor: COUI captures most of the gain, while UOCI shows negligible or negative improvement on several benchmarks (e.g., HeH$^+$) because retargeting cannot recover a trajectory already corrupted by uncontrolled noise during optimization. The full \sol{} combines both phases and consistently outperforms either partial configuration.

\begin{table}[t]
\centering
\caption{\textit{\sol{} shows lower standard deviation (\%) of performance evaluation metric (defined in Sec.~\ref{sec:methodology}).}}
\label{tab:performance_improvement_std}
\scalebox{0.73}{
\begin{tabular}{llcccccc}
\toprule
\rowcolor{headergray}
VQA & Case & \sol{} & Greedy & RandMAP & COUI & UOCI & Baseline \\
\midrule
\rowcolor{lightblue}
\textbf{VQE} & $H_2$ & 6.1\% & 9.2\% & 10.4\% & 12.3\% & 13.5\% & 16.2\% \\
\rowcolor{lightblue}
 & $HeH^+$ & 5.0\% & 7.3\% & 12.0\% & 9.1\% & 10.2\% & 13.4\% \\
\rowcolor{lightblue}
 & $H_3^+$ & 4.3\% & 6.2\% & 9.1\% & 11.4\% & 12.5\% & 15.4\% \\
\midrule
\rowcolor{lightgreen}
\textbf{QAOA} & Cycle & 5.1\% & 7.2\% & 8.4\% & 10.1\% & 12.3\% & 15.0\% \\
\rowcolor{lightgreen}
 & Ladder & 4.2\% & 6.3\% & 9.1\% & 11.2\% & 11.4\% & 15.6\% \\
\rowcolor{lightgreen}
 & Star & 3.8\% & 7.1\% & 8.5\% & 10.3\% & 12.1\% & 14.6\% \\
\midrule
\rowcolor{lightpink}
\textbf{QNN} & MNIST & 4.1\% & 7.0\% & 9.2\% & 10.5\% & 12.3\% & 14.8\% \\
\rowcolor{lightpink}
 & FashionMNIST & 3.7\% & 6.2\% & 8.3\% & 11.2\% & 12.1\% & 15.3\% \\
\bottomrule
\end{tabular}
}
\vspace{-3mm}
\end{table}

Stability is a central design goal of \sol{}. Each optimization-inference pair encounters different device conditions over the 30-day period, so variance in the performance metric reveals how fragile a technique's gains are. \sol{} achieves 3.7 -- 6.1\% standard deviation across all benchmarks, compared to 13.4 -- 16.2\% for the baseline (Table~\ref{tab:performance_improvement_std}). Consistent TVD and suppressed unreliable gradients (Sec.~\ref{sec:control_abstraction}) make outcomes depend on the control plane's decisions and not on which calibration window a run lands in.

\begin{figure}[t]
\centering
\includegraphics[scale=.29]{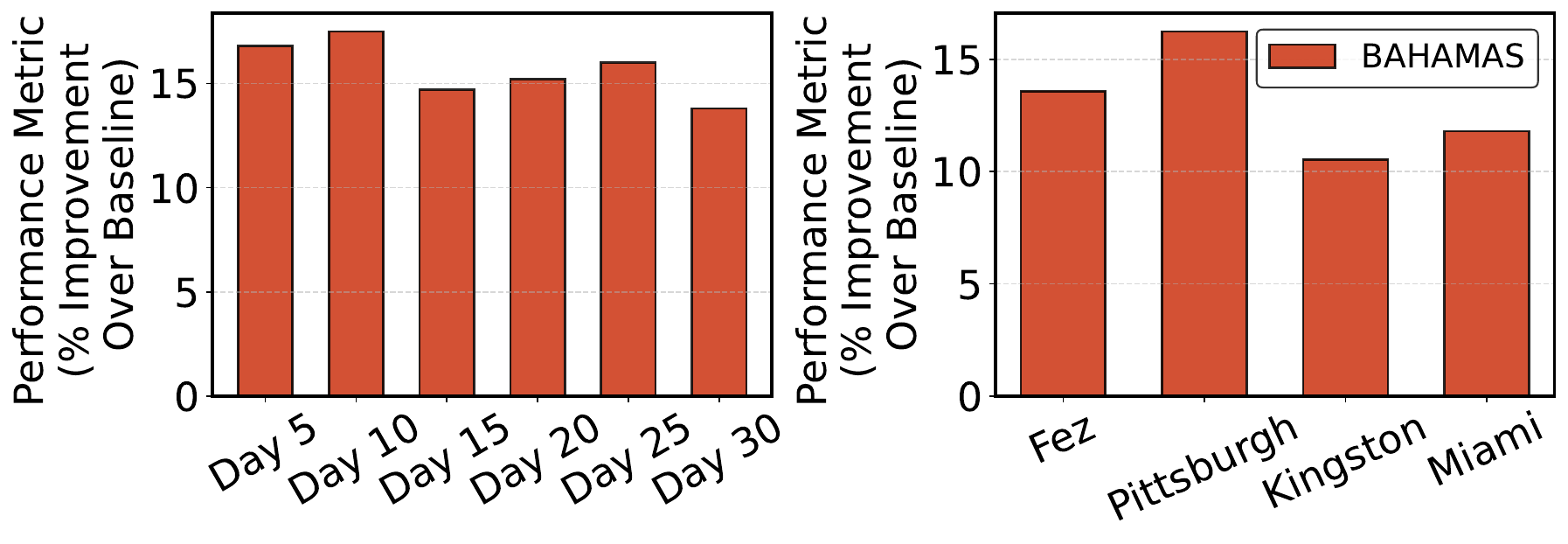}
\vspace{-5mm}
\caption{\textit{Inference-time retargeting -- optimized on IBM Pittsburgh (Day 1) and deployed on the same noise-drifted backend over 30 days (left) and on different backends (right).}}
\label{fig:inference}
\vspace{-4mm}
\end{figure}

We evaluate inference-time retargeting (Sec.~\ref{sec:inference}) by optimizing on IBM Pittsburgh and deploying under changed conditions, shown for VQE (H$_2$) in Fig.~\ref{fig:inference} (other benchmarks exhibit similar trends). On the same backend (left), \sol{} sustains 14 -- 17\% improvement over 30 days by matching estimated TVD to $\bar{\tau}_{\mathrm{opt}}$. Across backends (right), \sol{} maintains more than 11\% improvement on IBM Fez, Kingston, and Miami by jointly matching estimated TVD and the coupling structure. Performance varies with how well the execution signature matches the target hardware, but the consistency confirms the benefits of inference-time retargeting.

\begin{figure}[t]
\centering
\includegraphics[scale=.28]{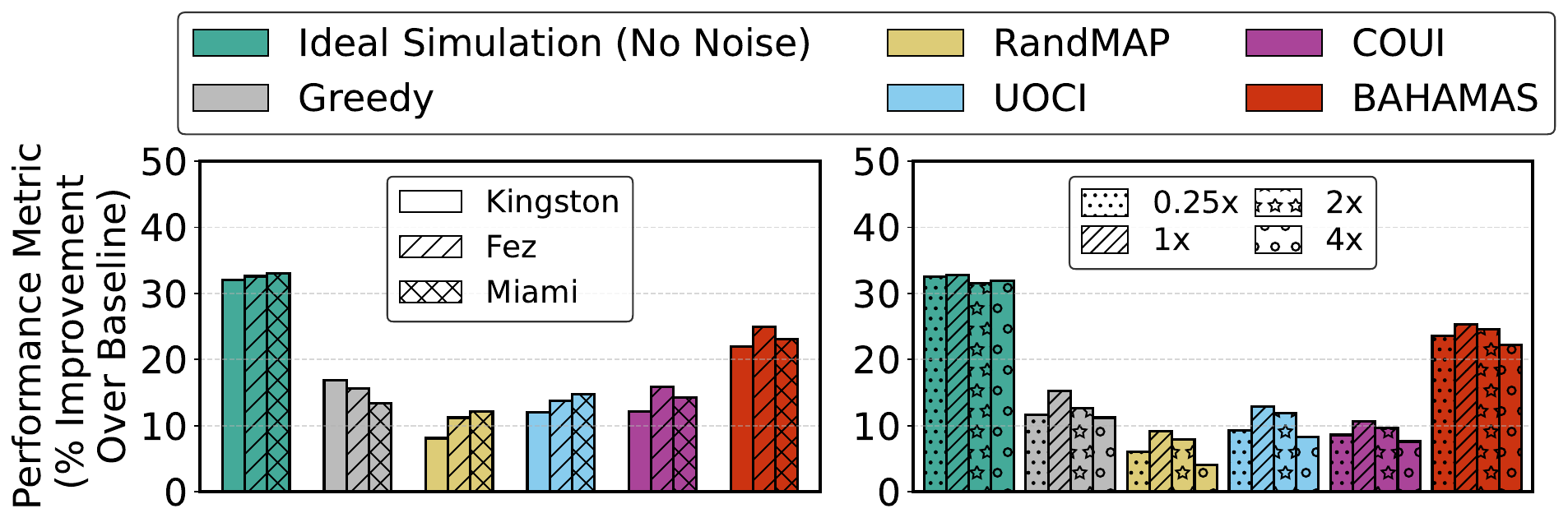}
\vspace{-4mm}
\caption{\textit{\sol{} delivers high performance across different quantum backends and noise levels.}}
\label{fig:noise_backends}
\vspace{-6mm}
\end{figure}

Fig.~\ref{fig:noise_backends} extends this to backend variation and noise scaling for QNN (FashionMNIST), with optimization and inference on the same backend. Across IBM Kingston, Fez, and Miami (left), \sol{} achieves 22--26\% improvement despite differences in processor generation and topology. On the right, noise on IBM Pittsburgh is scaled from 0.25$\times$ to 4$\times$ of the default using calibration-imported noisy execution (Sec.~\ref{sec:methodology}). At low noise, the baseline already performs reasonably. At moderate noise (1--2$\times$), \sol{}' control plane has the most leverage. At extreme noise, conditions shift faster than any reactive control can track, but \sol{} still retains over 20\% while RandMAP and COUI fall below 10\%.

\begin{figure}[t] 
\centering
\includegraphics[scale = .28]{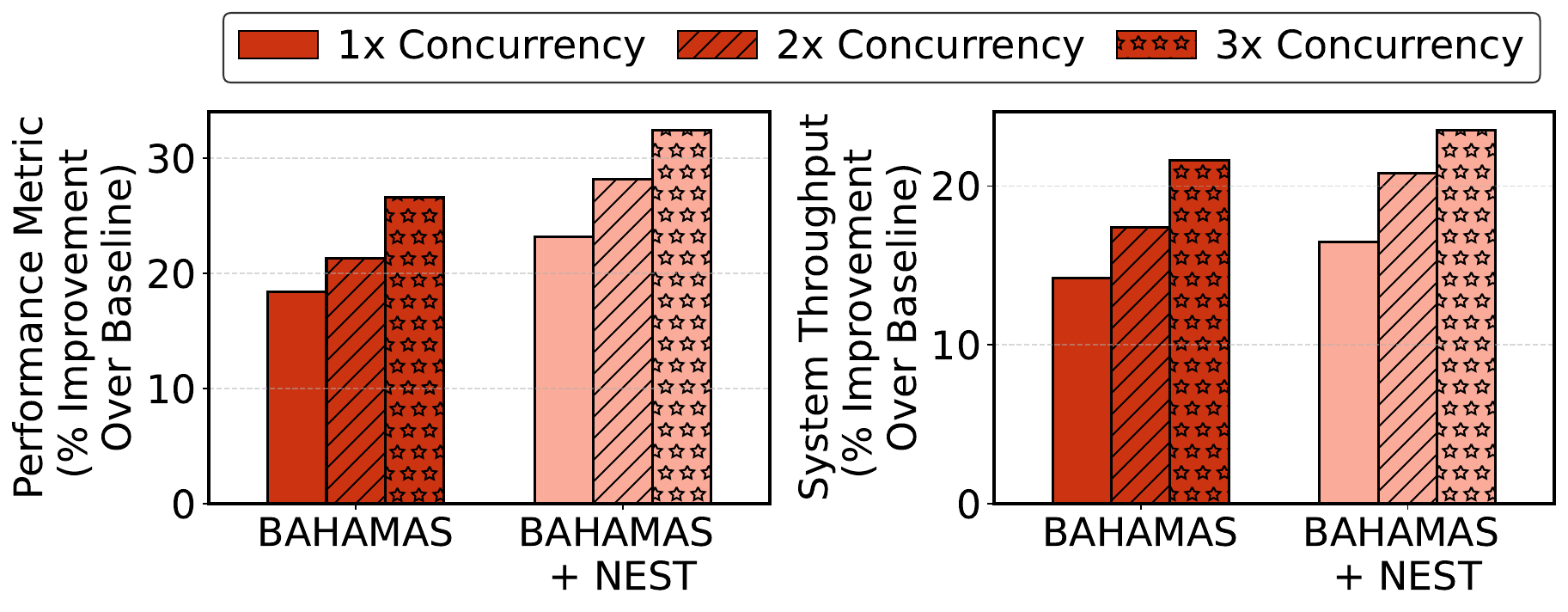}
\caption{\textit{\sol{} improves performance and throughput under concurrency for VQE when combined with NEST~\cite{huo2025three}.}}
\label{fig:concurrency}
\vspace{-3mm}
\end{figure}

To assess how \sol{} interacts with complementary techniques, we combine it with a recent work, NEST~\cite{huo2025three}, under increasing concurrency (1-3$\times$ co-located jobs) for VQE (H$_2$) in Fig.~\ref{fig:concurrency}. NEST varies qubit-fidelity regions across iterations and co-locates multiple jobs within non-overlapping chip regions to increase throughput. \sol{} operates within the assigned region and manages TVD stability and trust. Combined with NEST, \sol{} exceeds 30\% performance improvement at 3$\times$ concurrency. At higher concurrency, each job is confined to a smaller chip region, which constrains the candidate map space and makes within-region map selection more critical. \sol{}' per-iteration control ensures that even within a constrained region, the selected map maintains stable TVD and unreliable gradients are suppressed.

\vspace{2mm}

\noindent\textbf{Comparison with noise-adaptive execution (QISMET, DISQ).}
\textcolor{black}{We compare \sol{} against QISMET~\cite{ravi2023navigating} and DISQ~\cite{zhang2023disq}, prior techniques that also target noise-induced instability but keep the physical map fixed and react through temporal reruns or iteration skipping (Sec.~\ref{sec:methodology}). Averaged across all three VQA families (Table~\ref{tab:baseline_compare}), \sol{} reaches greater improvement over the uncontrolled baseline than QISMET (34.6\% versus 21\%) while converging in far fewer iterations (180 versus 1250), because co-located probing adds only a small fraction of per-iteration shots and recovered skips offset it, whereas QISMET reruns a full prior iteration at every step. \sol{} also preserves ranking better (Kendall $\tau$ of 0.83 versus 0.52, inversion probability of 15\% versus 24\%). Because DISQ is a fixed-map iteration-skipping method, it composes with our solution. \sol{}+DISQ raises the improvement of the performance evaluation metric to 48.4\%.}

\begin{table}[t]
\centering
\caption{\textit{\textcolor{black}{\sol{} versus prior noise-adaptive execution techniques, averaged across VQA families. \sol{} attains higher performance in fewer iterations with better ranking preservation, and composes with DISQ.}}}
\label{tab:baseline_compare}
\scalebox{0.77}{
\begin{tabular}{lccccc}
\toprule
\rowcolor{headergray}
Method & \makecell{Perf.\\(\% over base)} & \makecell{Iters. to\\converge} & \makecell{Shot\\overhead ($\times$)} & \makecell{Kendall\\$\tau$} & \makecell{Inversion\\prob.} \\
\midrule
QISMET~\cite{ravi2023navigating} & 21.0 & 1250 & 2.30 & 0.52 & 24\% \\
\sol{} & 34.6 & 180 & 1.93 & 0.83 & 15\% \\
\sol{}+DISQ~\cite{zhang2023disq} & 48.4 & 200 & 2.13 & 0.84 & 18\% \\
\bottomrule
\end{tabular}}
\vspace{-4mm}
\end{table}

\subsection{Reasons Behind \sol{}' Effectiveness}
\label{sec:eval_reason}

\textcolor{black}{\sol{}' gains come from preserving the relative ranking of parameter configurations across iterations, the property Observation~1 identifies as the root cause of destabilized optimization. To measure this directly, we compare each technique's on-hardware configuration ranking against the noiseless ranking using three metrics -- Kendall $\tau$ rank correlation, the probability of a pairwise ranking inversion, and top-$k$ stability ($k{=}3$). Averaged across all three VQA families (Table~\ref{tab:ranking}), \sol{} raises Kendall $\tau$ from 0.43 to 0.83, cuts inversion probability from 33\% to 15\%, and improves top-3 stability from 54\% to 88\%. Stable per-iteration TVD (Sec.~\ref{sec:control_abstraction}) holds the noise distortion on the cost constant, which is what preserves the ordering. These ranking metrics track the performance metric -- across runs, performance rises as Kendall $\tau$ rises and as inversion probability falls, with a Pearson correlation of $r\!=\!-0.79$ ($p<0.01$) between inversion probability and performance. Rank stability is therefore the cause of the gains, and it is not a side effect.}

\begin{table}[t]
\centering
\caption{\textit{\textcolor{black}{Ranking preservation against the noiseless configuration ranking, averaged across VQA families. \sol{} improves all the three metrics.}}}
\label{tab:ranking}
\footnotesize
\setlength{\tabcolsep}{4pt}
\begin{tabularx}{\columnwidth}{@{}Xcc@{}}
\toprule
\rowcolor{headergray}
Metric & Baseline & \sol{} \\
\midrule
Kendall $\tau$ versus noiseless ranking & 0.43 & 0.83 \\
Pairwise inversion probability & 33\% & 15\% \\
Top-$k$ stability ($k{=}3$) & 54\% & 88\% \\
\bottomrule
\end{tabularx}
\vspace{-2mm}
\end{table}

\begin{figure}[t] 
\centering
\includegraphics[scale = .29]{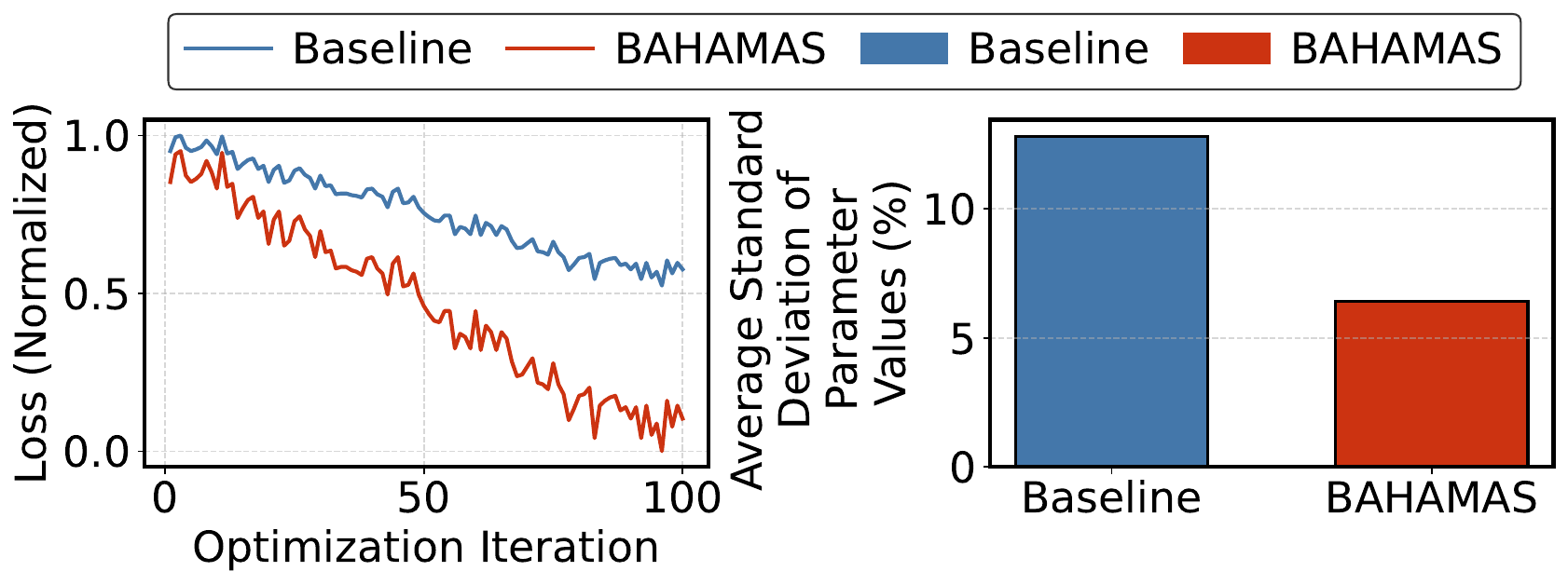}
\caption{\textit{\sol{} reduces loss over iterations and standard deviation of ansatz parameter values across different runs.}}
\label{fig:normalized_loss_std_plot}
\vspace{-6mm}
\end{figure}

The effectiveness of \sol{} also stems from keeping the optimization trajectory on track under hardware noise. The baseline loss (Fig.~\ref{fig:normalized_loss_std_plot}, left) plateaus above 0.5 after 100 iterations, while \sol{} reaches lower values in fewer iterations. Stable map selection and trust gating ensure only gradients computed under consistent noise are committed, preventing corrupted updates from accumulating. This is reflected in parameter variability (Fig.~\ref{fig:normalized_loss_std_plot}, right), the standard deviation of each ansatz parameter's final value across 100 independent QNN (MNIST) optimization runs. \sol{}'s standard deviation is much lower than the baseline, meaning the optimizer converges to similar values regardless of when the run executes, showing \sol{}'s stable optimization.

\begin{figure}[t] 
\centering
\includegraphics[scale = .29]{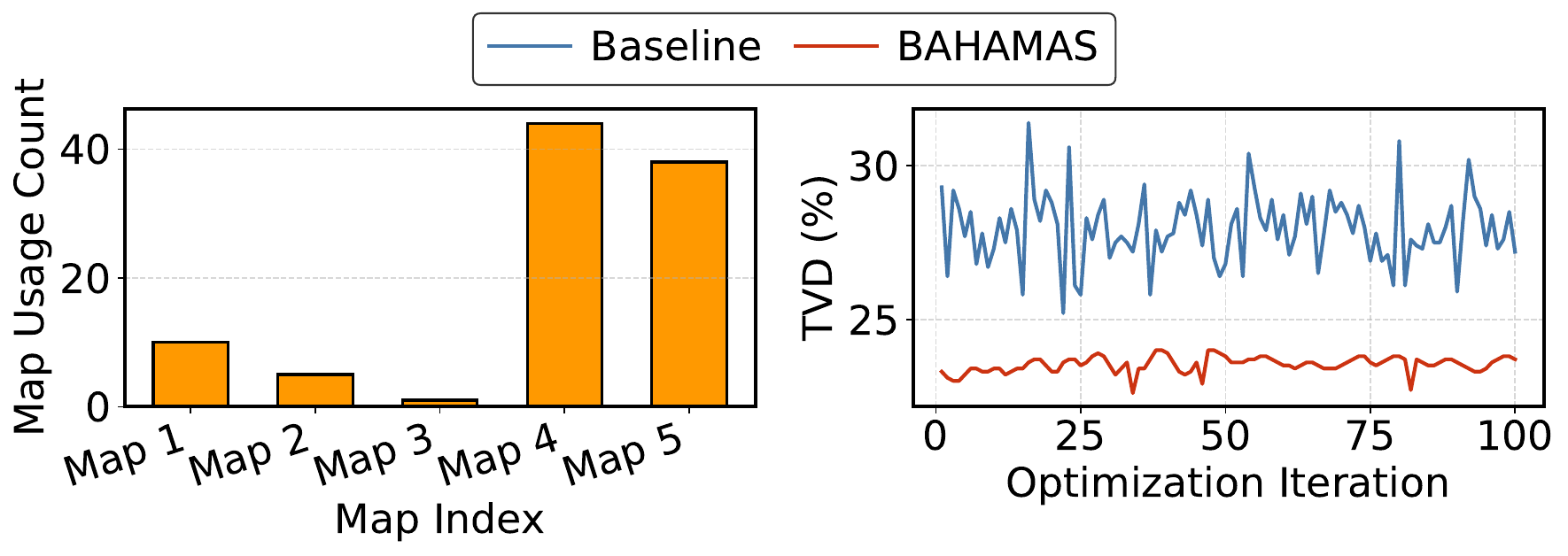}
\caption{\textit{\sol{} selects different maps to maintain lower TVD fluctuations, which helps in stabilizing optimization.}}
\label{fig:tvd_topo_plot}
\vspace{-4mm}
\end{figure}

\begin{figure}[t] 
\centering
\includegraphics[scale = .29]{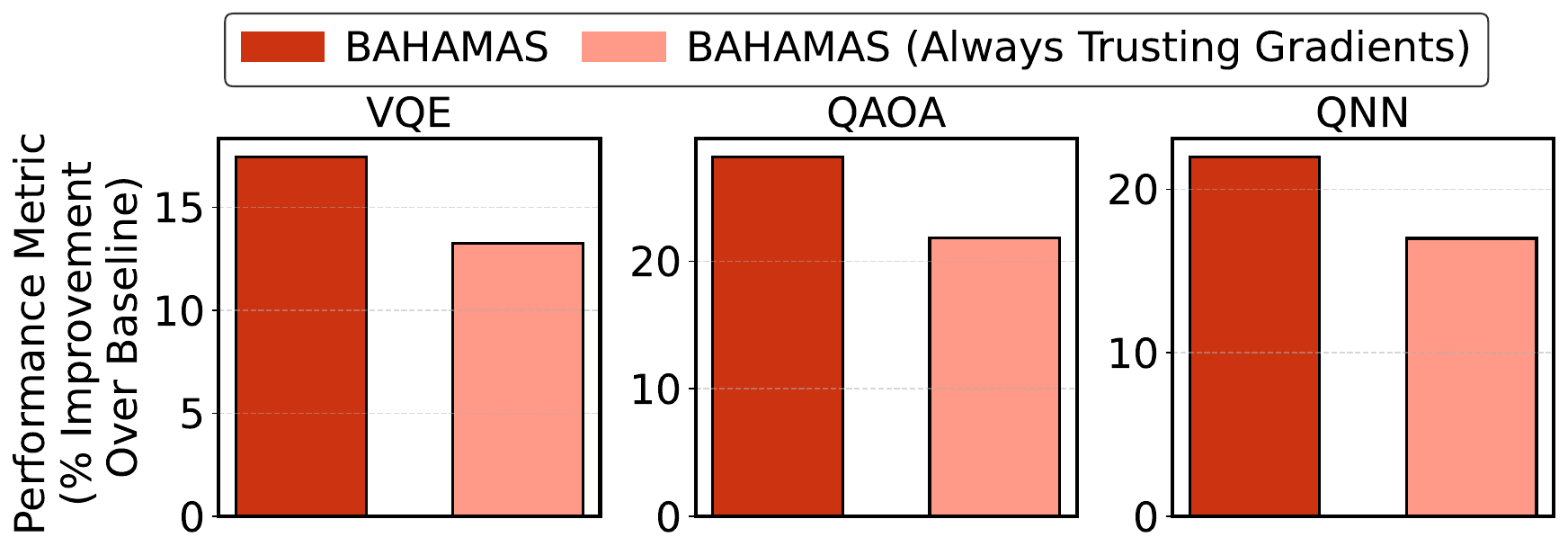}
\caption{\textit{Disabling \sol{}' trust gating and always committing gradients deteriorates performance.}}
\label{fig:bahamas_gradients}
\vspace{-6mm}
\end{figure}

Two key elements of \sol{}' design are TVD-stable map selection and conditional gradient trust. Fig.~\ref{fig:tvd_topo_plot} (left) shows that \sol{} distributes execution across multiple maps as device conditions shift (for VQE (H$_2$)). Different maps win at different iterations, which shows that the TVD estimator can track noise drift and update its rankings accordingly. The transpile-time estimator and execution-time co-location work together to keep the circuit in low-TVD regions. The effect (right) is both lower overall TVD and reduced fluctuation across iterations compared to the baseline. Lower fluctuations mean the noise distortion on the cost function remains consistent between consecutive parameter updates. This is critical because parameter-shift gradients compare paired circuit evaluations, and if the noise changes between them or between successive iterations, the resulting gradient direction is unreliable. By holding TVD stable, \sol{} helps the optimizer receive gradients that point in a consistent direction. Fig.~\ref{fig:bahamas_gradients} isolates trust gating by comparing \sol{} against a variant that always commits gradients. Disabling trust reduces improvement. Stable map selection improves the execution environment, but some iterations still encounter unfavorable noise. Without trust gating, corrupted gradients from those iterations are passed to the optimizer, undoing prior progress. Inference-time retargeting, the third key element of \sol{}, was evaluated in Fig.~\ref{fig:inference}.

The trust gating component specifically exploits gradient-based optimization, where each gradient ties to a known pair of parameter-shift evaluations under a specific map. Methods like SPSA, which estimate gradients from random perturbations, compound with hardware measurement noise and make the gradient signal inherently noisier than parameter-shift, regardless of device quality~\cite{bonet2023performance}. For example, SPSA underperforms even the uncontrolled gradient-based baseline by over 48 percentage points on QNN (MNIST) and 21 percentage points on QAOA (Cycle), with 2.2 -- 2.9$\times$ higher variance in both cases. The remaining key design components of \sol{} (TVD stabilization, map selection, consensus) are beneficial for even optimizers that are gradient-free.

\vspace{2mm}

\noindent\textbf{Parameter sensitivity.}
\textcolor{black}{\sol{} exposes four control parameters -- the target-TVD smoothing rate $\alpha$, the stability tolerance $\epsilon$, the consensus gate $\gamma_{\min}$, and the initial target TVD. We sweep each over the ranges in Table~\ref{tab:sensitivity} while holding the others at their defaults, on VQE (H$_2$). Performance varies by less than 3\% across every sweep, which means that the gains do not depend on fine-tuning. The exponential trust weight and the per-resource estimator keep the control loop stable across a wide operating band, and each parameter can be set from a simple heuristic (for example, the initial target from a percentile of the first probe set).}

\begin{table}[t]
\centering
\caption{\textit{\textcolor{black}{\sol{}' Performance stays within a few percentage points across the different parameter sweeps.}}}
\label{tab:sensitivity}
\footnotesize
\setlength{\tabcolsep}{4pt}
\begin{tabularx}{\columnwidth}{@{}XYc@{}}
\toprule
\rowcolor{headergray}
Parameter & Range tested & Performance\ variation \\
\midrule
$\alpha$ (target EMA) & 0.70 -- 0.95 & $<2\%$ \\
$\epsilon$ (stability tolerance) & 0.5$\times$--2$\times$ default & $<3\%$ \\
$\gamma_{\min}$ (consensus gate) & 0.5$\times$--2$\times$ default & $<2\%$ \\
Initial target TVD & 25th -- 75th percentile & $<2\%$ \\
\bottomrule
\end{tabularx}
\vspace{-4mm}
\end{table}

\subsection{Scalability and Robustness of \sol{}}

\begin{figure}[t] 
\centering
\includegraphics[scale = .29]{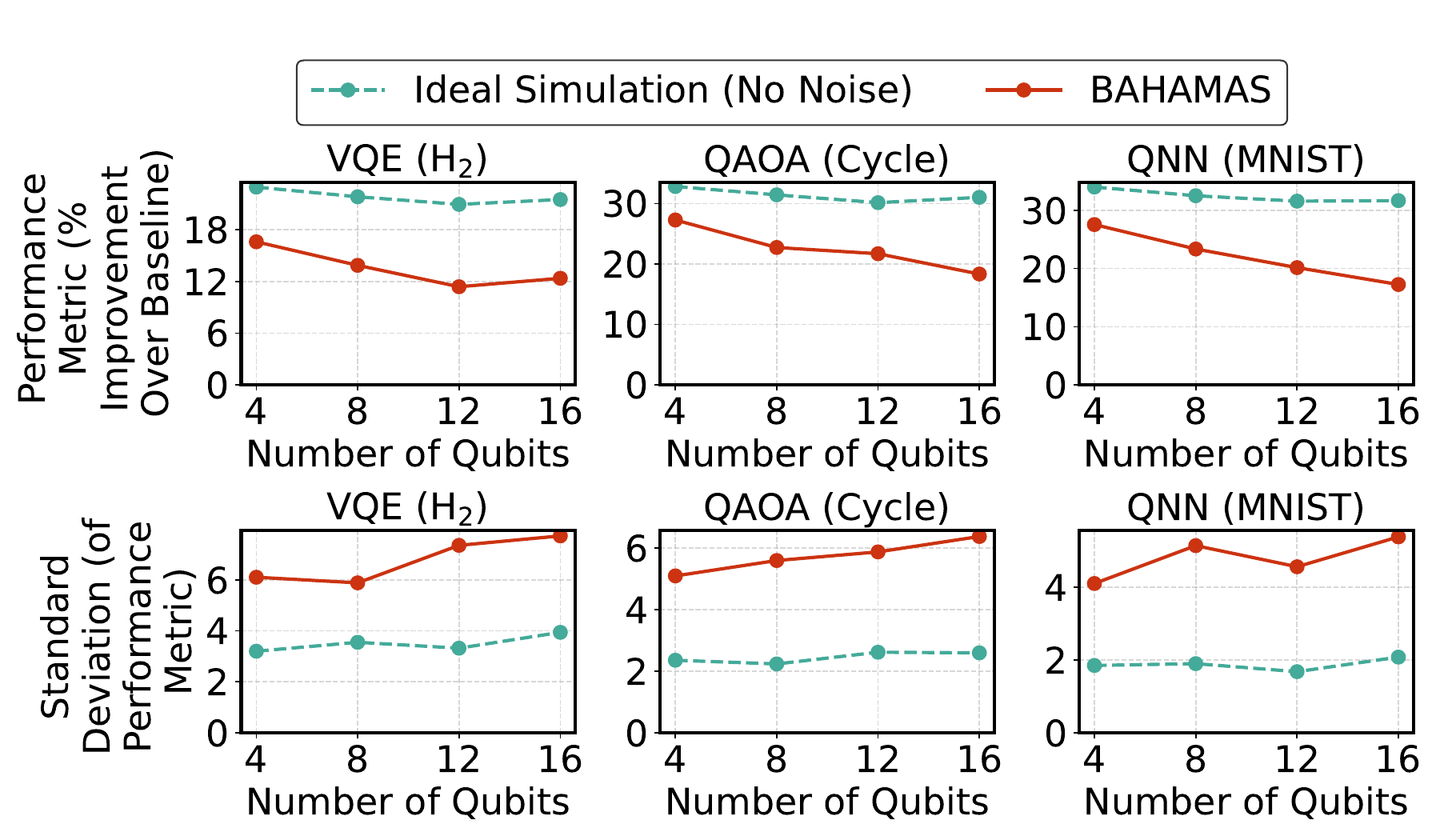}
\caption{\textit{\sol{} remains effective as ansatz size increases.}}
\label{fig:scaling}
\vspace{-6mm}
\end{figure}

We evaluate the robustness of \sol{} by scaling the number of qubits of each ansatz from 4 to 16 (Fig.~\ref{fig:scaling}). Larger qubit counts mean the routed circuit spans a wider region of the chip and passes through more physical gates and couplers, which increases overall noise exposure. \sol{} maintains significant improvement over the baseline at all qubit counts, retaining 12-19\% improvement even at 16 qubits. The gap to the ideal noiseless bound widens slightly because the accumulated noise from additional resources becomes harder to control. \sol{}' control plane remains effective -- it stabilizes the map-dependent TVD even as inherent noise from the larger footprint grows. Standard deviation (bottom row) across different optimization-inference runs stays between 4-7\% across all sizes, well below the baseline variance in Table~\ref{tab:performance_improvement_std}. \textcolor{black}{\sol{}' consensus operates over the small probe set, so the mechanism carries over directly as larger, lower-noise superconducting quantum devices become available. Larger backends also expose more candidate maps, which further increases the chances of finding stable operating regions.}

\begin{figure}[t] 
\centering
\includegraphics[scale = .29]{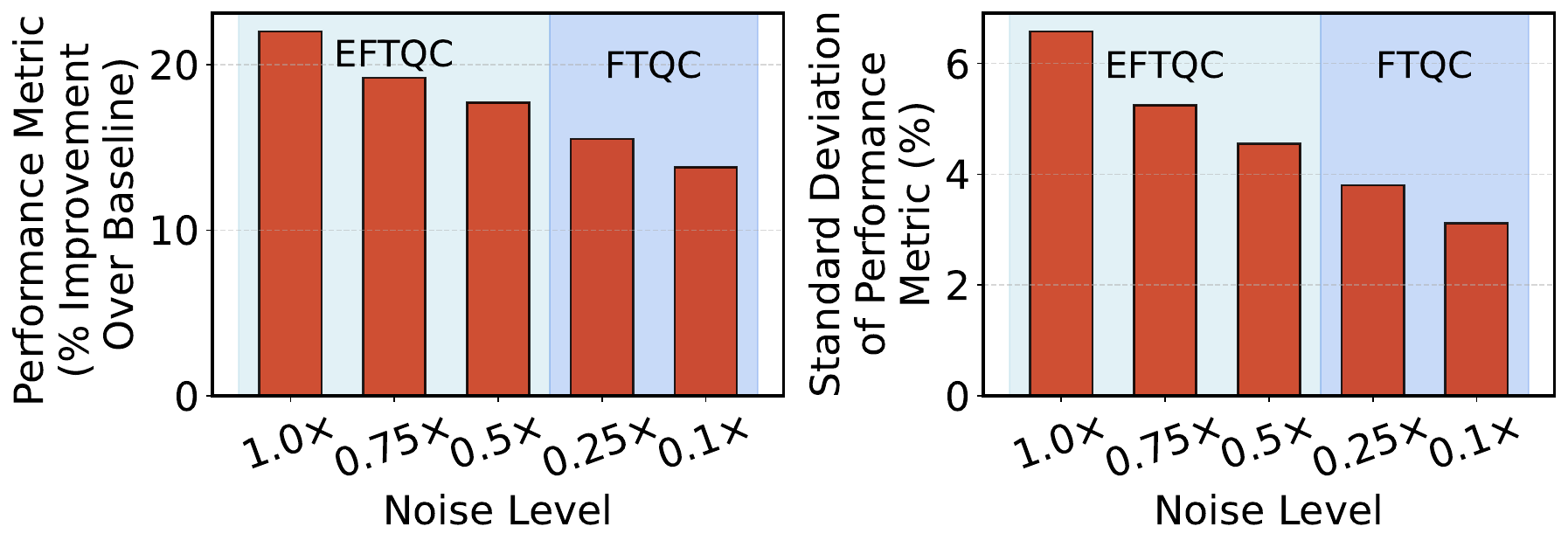}
\vspace{-4mm}
\caption{\textit{\sol{} is effective as we go toward FTQC-era.}}
\label{fig:noise_scaling_combined}
\vspace{-6mm}
\end{figure}

A natural question is whether \sol{} remains relevant as hardware improves. We test on the IBM Miami (early fault-tolerant (EFTQC)) Nighthawk processor and scale noise down to 0.1$\times$ to simulate fault-tolerant (FTQC) conditions (Sec.~\ref{sec:methodology}). At current noise (1.0$\times$), \sol{} achieves roughly 22\% improvement over the uncontrolled baseline (for VQE (H$_2$)), where map quality differences are largest (Fig.~\ref{fig:noise_scaling_combined}). As noise decreases, all maps perform closer to ideal, and the margin shrinks, but at 0.1$\times$ \sol{} still delivers 14\% improvement because qubit drift, calibration shifts, and temporal variation persist regardless of gate error rates. Standard deviation drops from 6.5\% to 3\%. This shows that \sol{}' control plane remains effective from EFTQC to FTQC era.


\section{Related Works}

\noindent\textbf{Noise mitigation and VQA trainability.}
Hardware noise limits VQA trainability through barren plateaus~\cite{wang2021noise} and corrupted signals. \textcolor{black}{Noise also reshapes the relative quality of parameter configurations, which in turn degrades the performance of variational algorithms, as in quantum chemistry~\cite{saib2021effect}.} Error mitigation~\cite{kim2023scalable,van2023probabilistic,czarnik2021error,kim2023evidence} recovers expectation values but depends on noise models that degrade under drift~\cite{kim2025error}, incurs exponential overhead~\cite{wang2024can}, and remains insufficient in early fault-tolerant regimes~\cite{dangwal2025variational}. System-level approaches~\cite{ravi2023navigating,dangwal2023varsaw,seifert2024clapton} handle individual noisy iterations but do not maintain noise consistency across the optimization loop. \textcolor{black}{For example, QISMET~\cite{ravi2023navigating} reruns prior iterations and DISQ~\cite{zhang2023disq} skips unreliable iterations, both on a single fixed physical map. \sol{} instead controls per-iteration map selection and is complementary to iteration skipping.} Algorithmic improvements~\cite{pesah2021absence,cerezo2021cost,grimsley2019adaptive,ravi2022cafqa,tang2021cutqc,ayanzadeh2023frozenqubits} address ansatz or problem structure, and recent work, NEST~\cite{huo2025three} schedules fidelity progression and co-locates jobs for throughput. None coordinates qubit assignment and parameter update quality per iteration across both optimization and inference.

\vspace{1mm}

\noindent\textbf{Noise-aware mapping and output estimation.}
Noise-aware mapping~\cite{tannu2019ensemble,tannu2019not,zulehner2019compiling,patel2020ureqa,patel2020veritas,9251970,li2019tackling,jin2024tetris} with extensions for crosstalk~\cite{murali2020software,xie2021mitigating} and gate reduction~\cite{jin2024tetris,xu2025optimizing} treats mapping as a one-time decision. Ensemble methods~\cite{tannu2019ensemble,patel2020veritas,ravi2022quancorde} diversify errors but do not adapt per iteration. ESP-based ranking~\cite{patel2023graphine,li2022optimal,brandhofer2023optimal,tannu2019ensemble,xie2021mitigating,patel2022geyser} misses circuit-level noise interactions~\cite{huo2025anchor}, and learning-based predictors~\cite{patel2021qraft,liu2020reliability,wang2022quest} require retraining as backends drift. Multi-programming~\cite{das2019case,liu2021qucloud} packs independent circuits for throughput but does not use co-location for quality estimation across candidate mappings. These methods rely on an external reference for output quality, either a noiseless simulator~\cite{patel2020veritas} or an offline-trained model~\cite{wang2022quest}, and neither holds as circuits scale and device noise drifts between iterations. BAHAMAS instead derives its quality signal from agreement among co-located maps at run time, and reuses that signal at inference.

\vspace{1mm}

\section{Conclusion}
\sol{} is an online control plane that stabilizes noise exposure across VQA iterations through per-resource TVD estimation, consensus-based quality signals from co-located execution, and trust-aware parameter update gating. It maintains a stable execution regime across transpilation, quantum execution, and classical parameter update. Evaluation on real hardware across multiple processor generations shows over 20\% average improvement, composability with complementary techniques, and effectiveness under fault-tolerant conditions. \sol{} is released at: https://doi.org/10.5281/zenodo.21513003, to serve as a system software for controlling the execution of variational quantum workloads on superconducting processors.

\vspace{2mm}

\noindent{\textbf{Acknowledgments.} We thank the reviewers for their constructive feedback. IBM Quantum resources were used for this work. All positions expressed in the paper are those of the authors and do not reflect the position of the IBM Quantum team. This research used the resources of the National Energy Research Scientific Computing Center, a DOE Office of Science User Facility supported by the Office of Science of the U.S. Department of Energy under Contract No. DE-AC02-05CH11231 using NERSC award NERSC DDR-ERCAP0037923. This work is supported by the University of Utah’s Kahlert School of Computing, Scientific Computing \& Imaging (SCI) Institute, and Rice University.}

\bibliographystyle{IEEEtranDOI}
\balance
\bibliography{paper_bahamas}
\end{document}